\documentclass[twocolumn,showpacs,preprintnumbers,amsmath,amssymb]{revtex4}

\usepackage{graphicx}
\usepackage{dcolumn}
\usepackage{bm}

\begin{document}

\preprint{APS/123-QED}

\title{White dwarf stars in $D$ dimensions}

\author{P.-H. Chavanis}
\email{chavanis@irsamc.ups-tlse.fr}
\affiliation{ Laboratoire de Physique Th\'eorique (UMR 5152 du CNRS), Universit\'e Paul Sabatier, 118
route de Narbonne 31062 Toulouse, France }%

\date{\today}

\begin{abstract}
We derive the mass-radius relation of relativistic white dwarf stars
(modeled as a self-gravitating degenerate Fermi gas at $T=0$) in a
$D$-dimensional universe and study the influence of the dimension of
space on the laws of physics when we combine quantum mechanics,
special relativity and gravity. We exhibit characteristic dimensions
$D=1$, $D=2$, $D=3$, $D=(3+\sqrt{17})/2$, $D=4$,
$D=2(1+\sqrt{2})$ and show that quantum mechanics cannot balance
gravitational collapse for $D\ge 4$. This is similar to a result found
by Ehrenfest (1917) at the atomic level for Coulomb forces (in Bohr's
model) and for the Kepler problem. This makes the dimension of our
universe $D=3$ very particular with possible implications regarding
the anthropic principle. We discuss some historic aspects concerning
the discovery of the Chandrasekhar (1931) limiting mass in relation to
previous investigations by Anderson (1929) and Stoner (1930). We also
propose different derivations of the stability limits of polytropic
distributions and consider their application to classical and relativistic
white dwarf stars.
\end{abstract}

\maketitle

\section{Introduction: some historic elements}
\label{sec_introduction}

The study of stars is one of the most fascinating topics in
astrophysics because it involves many areas of physics: gravitation,
thermodynamics, hydrodynamics, statistical mechanics, quantum
mechanics, relativity,... and it has furthermore a very interesting
history. In his classical monograph {\it The Internal Constitution of
the Stars}, Eddington (1926) \cite{eddington0} lays down the
foundations of the subject and describes in detail the basic processes
that govern the structure of ordinary stars. At that stage, only
elements of ``classical'' physics are used and difficulties with such
theories to account for the structure of high density stars such as
the companion of Sirius are pointed out.  Soon after the discovery of
the quantum statistics by Fermi (1926) \cite{fermi} and Dirac (1926)
\cite{dirac}, Fowler (1926) \cite{fowler} uses this ``new
thermodynamics'' to explain the puzzling nature of white dwarf
stars. He understands that low mass white dwarf stars owe their
stability to the quantum pressure of the degenerate electron gas 
\footnote{In Fowler's work, it is assumed that the system is
completely degenerate ($T=0$). The case of a partially degenerate
self-gravitating Fermi gas at arbitrary temperature has been discussed
more recently by Hertel \& Thirring (1971) \cite{ht} and Chavanis
(2002) \cite{pt} in the context of statistical mechanics. They
describe the phase transition, below a critical temperature $T_c$,
from a gaseous configuration to a condensed state (with a degenerate
core surrounded by a halo). This provides a simple physical mechanism
showing how the system can reach highly degenerate configurations as a
result of gravitational collapse; see Chavanis (2006) \cite{review}
for a review on phase transitions in self-gravitating systems.}. The
resulting structure is equivalent to a polytrope of index $n=3/2$ so
that the mass-radius relation of classical white dwarf stars behaves
like $MR^{3}\sim 1$ (Chandrasekhar 1931a
\cite{chandra0}). The next step was made by Chandrasekhar, aged only
nineteen, who was accepted by the University of Cambridge to work with
Fowler.  In the boat that took him from Madras to Southampton
\cite{wali}, Chandrasekhar understands that relativistic effects are
important in massive white dwarf stars and that Einstein kinematic
must be introduced in the problem. In his first treatment
(Chandrasekhar 1931b \cite{chandra1}), he considers the
ultra-relativistic limit and shows that the resulting structure is
equivalent to a polytrope of index $n=3$. Applying the theory of
polytropic gas spheres (Emden 1907 \cite{emden}), this leads to a
unique value of the mass $M_{c}$ that he interprets as a limiting
mass, nowdays called the Chandrasekhar limit \footnote{Landau (1932)
\cite{landau} made, independently, an equivalent calculation and
argued that for $M>M_{c}$ quantum mechanics cannot prevent the system
from collapsing to a point. However, he did not take this collapse
very seriously and noted that, in reality, there exists stars with
mass $M>M_{c}$ that do not show this ``ridiculous tendency'', so that
they must possess regions in which the laws of quantum mechanics are
violated.}.  The complete mass-radius relation of relativistic white
dwarf stars was given later (Chandrasekhar 1935 \cite{chandra3}) and
departs from Fowler's sequence as we approach the limiting mass. At
this critical mass, the radius of the configuration vanishes.  Above
the critical mass, the equation of state of the relativistic
degenerate Fermi gas of electrons is not able to balance gravitational
forces and, when considering the final evolution of such a star,
Chandrasekhar (1934)
\cite{chandra2} ``left speculating on other
possibility''. Chandrasekhar's result was severly criticized by
Eddington (1935) \cite{eddington} who viewed this result as {\it a
reductio ad absurdum} of the relativistic formula and considered the
combination of special relativity and non-relativistic quantum theory
as an ``unholly alliance''. Because of the querelle with Eddington, it
took some time to realize the physical implication of Chandrasekhar's
results. However, progressively, his early investigations on white
dwarf stars were extended in general relativity to the case of neutron
stars (Oppenheimer \& Volkoff 1939 \cite{ov}) and finally led to the
concept of ``black holes'', a term coined by Wheeler in 1967. Without
the dispute with Eddington, this ultimate stage of matter resulting
from gravitational collapse could have been predicted much earlier
from the discovery of Chandrasekhar
\cite{zakharov}.

Although these historic elements are well-known, it is less well-known
that the concept of a maximum mass for relativistic white dwarf stars
had been introduced earlier by Anderson (1929) \cite{anderson} and
Stoner (1930) \cite{stoner}
\footnote{This historical point has been noted, independently, by 
Blackman \cite{blackman}.}. These studies are mentioned in the early
works of Chandrasekhar but they have been progressively forgotten and
are rarely quoted in classical textbooks of astrophysics. These
authors investigated the equation of state of a relativistic
degenerate Fermi gas and predicted an upper limit for the mass of
white dwarf stars. Stoner (1930) \cite{stoner} uses a uniform mass
density to model the star while Chandrasekhar (1931) \cite{chandra1}
considers a more realistic $n=3$ polytrope. However, as noted in
Chandrasekhar (1931) \cite{chandra1}, the value of the limiting mass
found by Stoner with his simplified model is relatively close to that
obtained with the improved treatment. Interestingly, Nauenberg (1972)
\cite{nauenberg} introduced long after a simplified treatment of
relativistic white dwarf stars in order to obtain an analytical
approximation of the mass-radius relation. It turns out that this
model, which gives a very good agreement with Chandrasekhar's
numerical results (up to some normalization factors), is similar to
that introduced by Stoner.

Leaving aside these interesting historical remarks, the result of
Chandrasekhar \cite{chandra1} concerning the existence of a limiting
mass is very profound because this mass can be expressed in terms of
fundamental constants, similarly to the Bohr radius of the hydrogen
atom \cite{chandra4}. Hence, the mass of stars is determined typically
by the following combination
\begin{equation}
\left \lbrack {hc\over G}\right \rbrack^{3/2}{1\over H^{2}}\simeq 29.2 M_{\odot}
\end{equation}
where $G$ is the constant of gravity, $h$ the Planck constant, $c$ the
velocity of light and $H$ the mass of the hydrogen atom ($M_{\odot}$
is the solar mass). This formula results from the combination of
quantum mechanics ($h$), special relativity ($c$) and gravity
$(G)$. Since dimensional analysis plays a fundamental role in physics,
it is of interest to investigate how the preceding results depend on
the dimension of space $D$ of the universe. In our previous
investigations \cite{lang,fermiD,aaantonov}, we considered the case of
classical white dwarf stars in $D$ dimensions and found that they
become unstable in a space of dimension $D\ge 4$. In that case, the
star either evaporates or collapses. Therefore, quantum mechanics
cannot stabilize matter against gravitational collapse in $D\ge 4$
contrary to what happens in $D=3$
\cite{fowler,ht,pt}. Interestingly, this is similar to a result found by
Ehrenfest \cite{ehrenfest} at the atomic level for Coulomb forces in
Bohr's model and for the planetary motion (Kepler problem). The object
of this paper is to extend these results to the case of relativistic
white dwarf stars and exhibit particular dimensions of space which
play a special role in the problem when we combine Newtonian gravity,
quantum mechanics and special relativity.  We shall see that the
problem is very rich and interesting in its own right. It shows to
which extent the dimension $D=3$ of our universe is particular, with
possible implications regarding the anthropic principle
\cite{barrowBOOK}. We note that a similar problem has been considered
in \cite{chabrier} on the basis of dimensional analysis. Our approach
is more precise since we generalize the {\it exact} mathematical
treatment of Chandrasekhar \cite{chandra3} to a space of dimension
$D$. A connection with other works investigating the role played by
the dimension of space on the laws of physics is made in the
Conclusion. In Appendix \ref{sec_stab}, we propose different
derivations of the stability limits of polytropic spheres and consider
applications of these results to classical and relativistic white
dwarf stars.

\section{The equation of state}
\label{sec_es}

Following Chandrasekhar \cite{chandra3}, we model a white dwarf star
as a degenerate gas sphere in hydrostatic equilibrium. The pressure is
due to the quantum properties of the electrons and the density to the
protons. In the completely degenerate limit, the electrons have
momenta less than a threshold value $p_{0}$ (Fermi momentum) and their
distribution function is $f=2/h^{D}$ where $h$ is the Planck
constant. There can only be two electrons in a phase space element of
size $h^D$ on account of the Pauli exclusion principle.  Therefore,
the number of electrons per unit volume is
\begin{equation}
n=\int f d{\bf p}={2S_{D}\over h^{D}}\int_{0}^{p_{0}}p^{D-1}dp={2S_{D}\over D h^{D}}p_{0}^{D},
\label{es1}
\end{equation}
where $S_{D}=2\pi^{D/2}/\Gamma(D/2)$  is the surface of a unit sphere in $D$-dimensions. The mean kinetic energy per electron is given by
\begin{equation}
\kappa={1\over n}\int f \epsilon(p) d{\bf p}={2S_{D}\over n h^{D}}\int_{0}^{p_{0}}\epsilon(p) p^{D-1}dp,
\label{es2}
\end{equation}
where $\epsilon(p)$ is the energy of an electron with impulse $p$. In relativistic mechanics,
\begin{equation}
\epsilon=mc^{2}\biggl\lbrace \biggl (1+{p^{2}\over m^{2}c^{2}}\biggr )^{1/2}-1\biggr\rbrace.
\label{es3}
\end{equation}
The pressure of the electrons is
\begin{equation}
P={1\over D}\int f p{d\epsilon\over dp}d{\bf p}={2S_{D}\over Dh^{D}}\int_{0}^{p_{0}}p^{D}{d\epsilon\over dp}dp.
\label{es4}
\end{equation}
Using Eq. (\ref{es3}), the pressure can be rewritten
\begin{equation}
P={2S_{D}\over Dmh^{D}}\int_{0}^{p_{0}}{p^{D+1}\over  (1+{p^{2}\over m^{2}c^{2}} )^{1/2}}dp.
\label{es5}
\end{equation}
Finally, the mass density of the star is
\begin{equation}
\rho=n\mu H,
\label{es6}
\end{equation}
where $H$ is the mass of the proton and $\mu$ the molecular
weight. If we consider a pure gas of fermions (like, e.g., massive
neutrinos in dark matter models), we just have to replace $\mu H$ by
their mass $m$.

Introducing the notation $x=p_{0}/mc$, we can write the density of
state parametrically as follows
\begin{equation}
P=A_{2}f(x), \qquad \rho=Bx^{D},
\label{es7}
\end{equation}
where
\begin{equation}
A_{2}={S_{D}m^{D+1}c^{D+2}\over 4Dh^{D}}, \qquad
B={2S_{D}m^{D}c^{D}\mu H\over Dh^{D}}, \label{es8}
\end{equation}
\begin{equation}
f(x)=8\int_{0}^{x}{t^{D+1}\over  (1+t^{2} )^{1/2}}dt.
\label{es9}
\end{equation}
The function $f(x)$ has the asymptotic behaviors
\begin{equation}
f(x)\simeq {8\over D+2}x^{D+2} \qquad (x\ll 1)
\label{es10}
\end{equation}
\begin{equation}
f(x)\simeq {8\over D+1}x^{D+1} \qquad (x\gg 1)
\label{es11}
\end{equation}
The classical limit corresponds to $x\ll 1$ and the ultra-relativistic
limit to $x\gg 1$. Explicit expressions of the function $f(x)$ are given in
Appendix \ref{sec_exp} for different dimensions of space.

\section{The Chandrasekhar equation}
\label{sec_c}

For a spherically symmetric distribution of matter, the equations of
hydrostatic equilibrum are
\begin{equation}
{dP\over dr}=-{GM(r)\over r^{D-1}}\rho,
\label{c1}
\end{equation}
\begin{equation}
M(r)=\int_{0}^{r}\rho S_{D}r^{D-1}dr.
\label{c2}
\end{equation}
They can be combined to give
\begin{equation}
{1\over r^{D-1}}{d\over dr}\biggl ({r^{D-1}\over \rho}{dP\over dr}\biggr )=-S_{D}G\rho.
\label{c3}
\end{equation}
Expressing $\rho$ and $P$ in terms of $x$ and setting $y^{2}=1+x^{2}$, we obtain
\begin{equation}
{1\over r^{D-1}}{d\over dr}\biggl ({r^{D-1}}{dy\over dr}\biggr )=-{S_{D}GB^{2}\over 8A_{2}}(y^{2}-1)^{D/2}.
\label{c4}
\end{equation}
We denote by $x_{0}$ and $y_{0}$ the values of $x$ and $y$ at the center. Furthermore, we define
\begin{equation}
r=a\eta, \qquad y=y_{0}\phi,
\label{c5}
\end{equation}
\begin{equation}
a=\biggl ({8A_{2}\over S_{D}G}\biggr )^{1/2}{1\over By_{0}^{{(D-1)/ 2}}}, \qquad y_{0}^{2}=1+x_{0}^{2}.
\label{c6}
\end{equation}
Note that the scale of length $a$ is independent on $y_0$ for $D=1$.
Substituting these transformations in Eq. (\ref{c4}), we obtain the
$D$-dimensional generalization of Chandrasekhar's differential
equation
\begin{equation}
{1\over \eta^{D-1}}{d\over d\eta}\biggl ({\eta^{D-1}}{d\phi\over d\eta}\biggr )=-\biggl (\phi^{2}-{1\over y_{0}^{2}}\biggr )^{D/2},
\label{c7}
\end{equation}
with the boundary conditions
\begin{equation}
\phi(0)=1, \qquad \phi'(0)=0.
\label{c8}
\end{equation}
The radius $R$ of the star is such that $\rho(R)=0$. This yields
\begin{equation}
\phi(\eta_{1})={1\over y_{0}}.
\label{c9}
\end{equation}
The density can be expressed as
\begin{equation}
\rho=\rho_{0}{y_{0}^{D}\over (y_{0}^{2}-1)^{D/2}}\biggl (\phi^{2}-{1\over y_{0}^{2}}\biggr )^{D/2},
\label{c10}
\end{equation}
where the central density
\begin{equation}
\rho_{0}=Bx_{0}^{D}=B(y_{0}^{2}-1)^{D/2}.
\label{c11}
\end{equation}
Finally, we find that the mass is related to $y_{0}$ by
\begin{equation}
M=-S_{D}\biggl ({8A_{2}\over S_{D}G}\biggr )^{D/2}{1\over B^{D-1}}y_{0}^{D(3-D)/2}\biggl (\eta^{D-1}{d\phi\over d\eta}\biggr )_{\eta=\eta_{1}}.
\label{c12}
\end{equation}
Note that $y_0$ does not {\it explicitly} enter in this expression
for $D=3$ but it is of course present implicitly.

\section{The classical limit}
\label{sec_class}

In the classical case $x\ll 1$, we find that the equation of state takes the form
\begin{equation}
P=K_{1}\rho^{1+{2/D}},
\label{class1}
\end{equation}
with
\begin{equation}
 K_{1}={1\over D+2}\biggl ({D\over 2S_{D}}\biggr )^{2/ D}{h^{2}\over m (\mu H)^{(D+2)/D}}.
\label{class2}
\end{equation}
Therefore a classical white dwarf star is equivalent to a polytrope of
index \cite{lang} \footnote{The lower script $3/2$ corresponds to the
value of the polytropic index of a classical white dwarf star in
$D=3$. The index $n_{3/2}$ refers to its corresponding value in $D$
dimensions. The same convention is adopted for the other indices $n_{3}$,
$n_{5}$ and $n_{3}'$.}:
\begin{equation}
n_{3/2}={D\over 2}.
\label{class3}
\end{equation}
Polytropic stars are described by the Lane-Emden equation
\cite{emden}. This can be recovered as a limit of the Chandrasekhar
equation. For $x\ll 1$, we have $y_{0}\simeq 1+{1\over
2}x_{0}^{2}$. We define $\theta=\phi^{2}-1/y_{0}^{2}$. To leading
order, $\phi=1-(x_{0}^{2}-\theta)/2$. Setting
$\xi=\sqrt{2}\eta$ and combining the foregoing results, we find that
Eq. (\ref{c7}) reduces to the Lane-Emden equation with index $D/2$:
\begin{equation}
{1\over\xi^{D-1}}{d\over d\xi}\biggl (\xi^{D-1}{d\theta\over d\xi}\biggr )=-\theta^{D/2},
\label{class4}
\end{equation}
\begin{equation}
\theta(0)=x_{0}^{2}, \qquad \theta'(0)=0.
\label{class5}
\end{equation}
Note that the condition at the origin is $\theta(0)=x_{0}^{2}$ instead
of $\theta(0)=1$ as in the ordinary Lane-Emden equation.  However,
using the homology theorem for polytropic spheres \cite{chandrabook}, we
can easily relate $\theta$ to $\theta_{D/2}$, the solution of the
Lane-Emden equation with index $n_{3/2}=D/2$ and condition at the
origin $\theta(0)=1$.

\begin{figure}[htbp]
\centerline{
\includegraphics[width=8cm,angle=0]{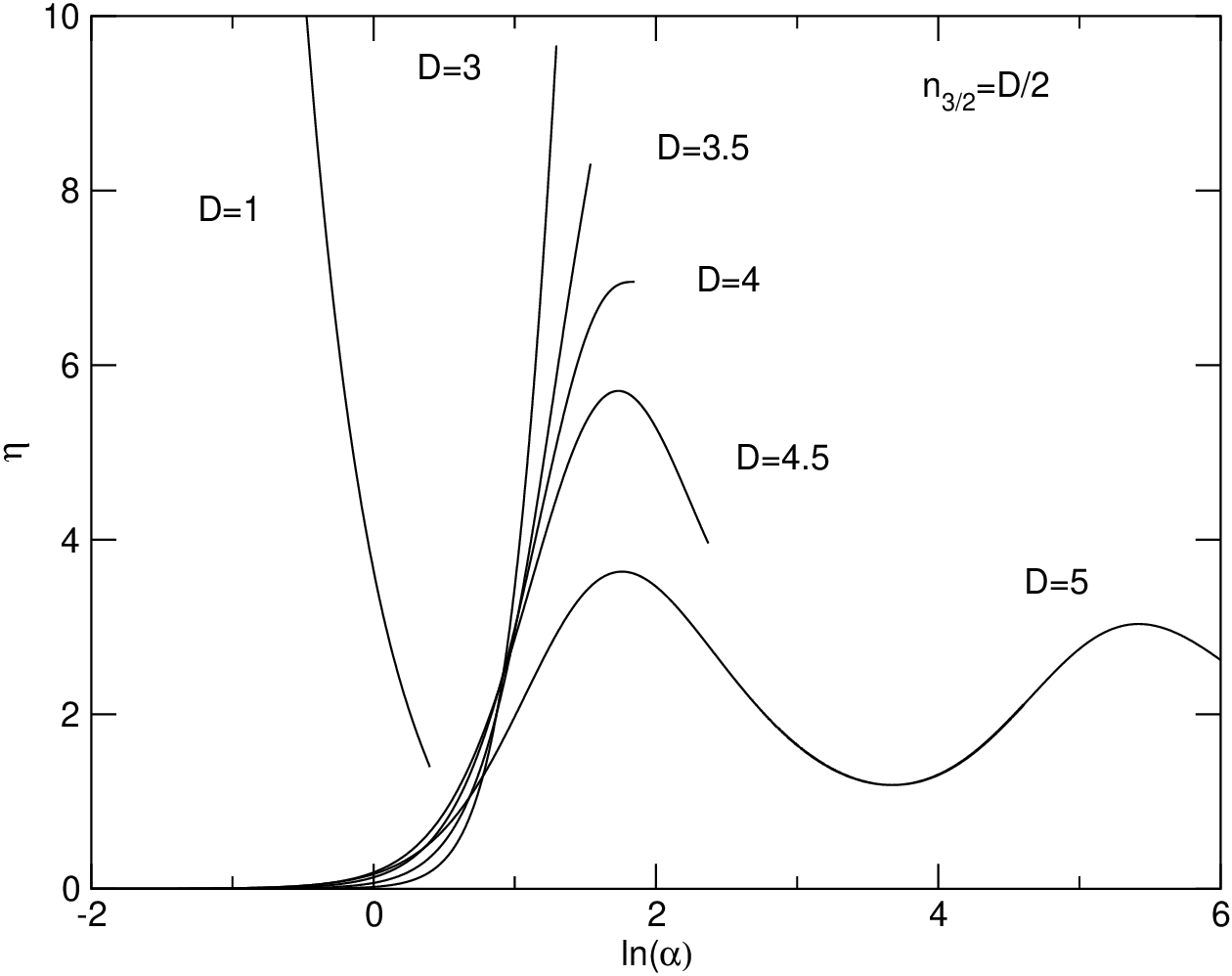}
} \caption[]{Relation between the mass (ordinate) and the central
density (abscissa) of box-confined polytropes with index $n_{3/2}=D/2$
in an appropriate system of coordinates (see \cite{lang,aaantonov} for
details). Complete polytropes correspond to the terminal point in the
series of equilibria. The series becomes dynamically unstable with
respect to the Euler-Poisson system (saddle point of the energy
functional) after the turning point of mass $\eta$ which appears for
$D=4$. Thus, classical white dwarf stars are stable for $D<4$ and
unstable for $D\ge 4$. }
\label{alphaC}
\end{figure}

\begin{figure}[htbp]
\centerline{
\includegraphics[width=8cm,angle=0]{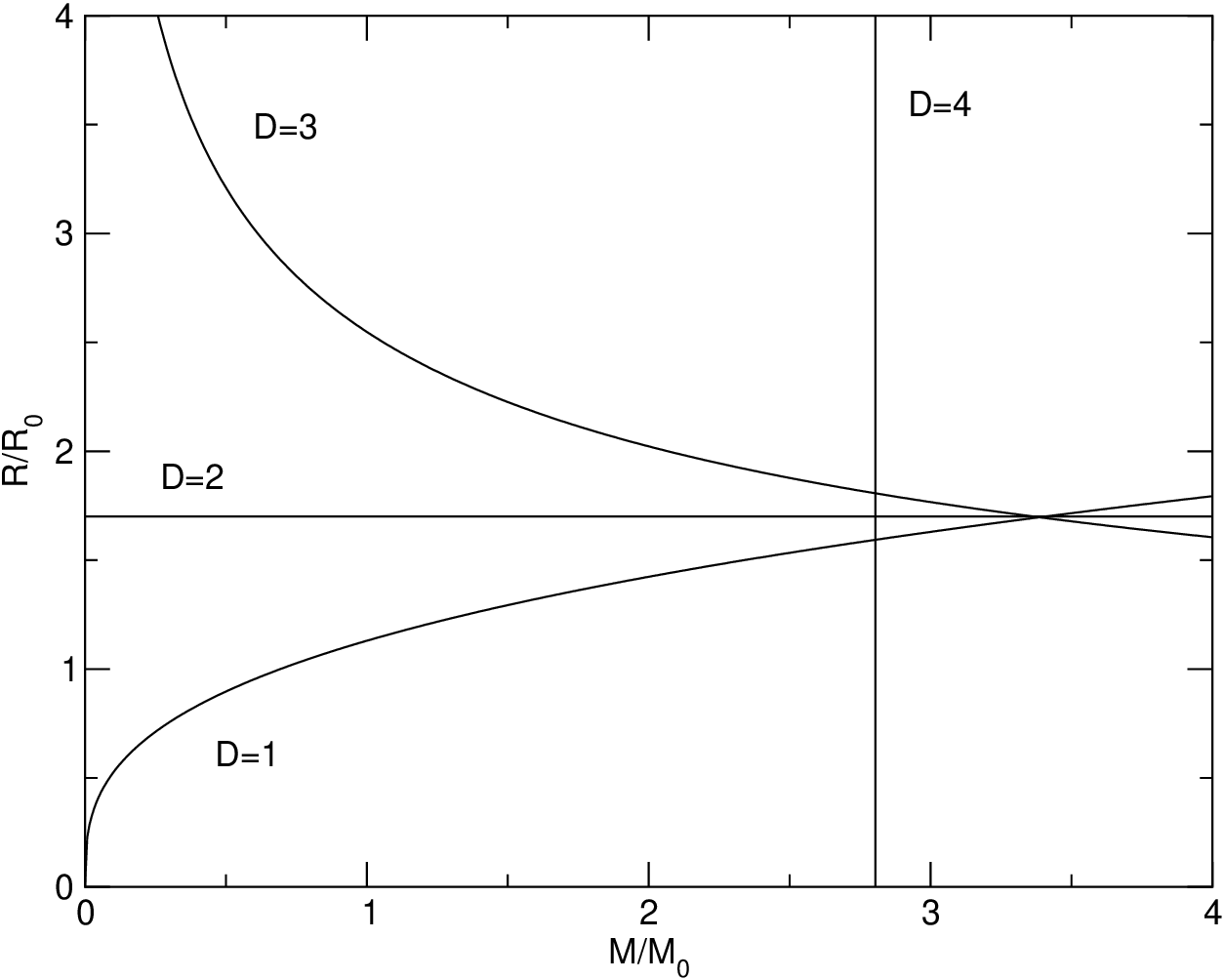}
} \caption[]{Mass-radius relation for classical white dwarf stars in
different dimensions of space. The radius is independent on mass in
$D=2$ and the mass is independent on radius in $D=4$. The dimensional factors $M_{0}$ and $R_{0}$ are defined in Sec. \ref{sec_gen}. }
\label{classique}
\end{figure}

The structure and the stability of polytropic spheres in various
dimensions of space has been studied by Chavanis \& Sire \cite{lang}.
For $D>2$, this study exhibits two important indices:
\begin{equation}
n_{3}={D\over D-2}, \qquad n_{5}={D+2\over D-2}.
\label{class6}
\end{equation}
According to this study, a classical white dwarf star is self-confined
(complete) if $n_{3/2}<n_{5}$, i.e. $D<2(1+\sqrt{2})=4.8284271...$. In
that case, the density vanishes at a finite radius $R$ identified
as the radius of the star.  On the other hand, it is nonlinearly
dynamically stable with respect to the Euler-Poisson system if
$n_{3/2}<n_{3}$, and linearly unstable otherwise. Therefore, classical white
dwarf stars are stable for $D<4$ and unstable for $D\ge 4$ (see
Fig. \ref{alphaC} and Appendix
\ref{sec_stab}). For $D\le 2$, classical white dwarf stars are always self-confined and stable. Using the results of \cite{lang}, the
mass-radius relation for complete polytropes with index $n_{3/2}=D/2$
in $D$ dimensions is
\begin{equation}
M^{(D-2)/D}R^{4-D}={K_{1}(D+2)\over 2GS_{D}^{2/D}}\omega_{D/2}^{(D-2)/D},
\label{class7}
\end{equation}
where we have defined
\begin{equation}
\omega_{D/2}=-\xi_{1}^{D+2\over D-2}\theta'_{D/2}(\xi_{1}),
\label{class8}
\end{equation}
where $\xi_{1}$ is such that $\theta_{D/2}(\xi_{1})=0$. Using
Eq. (\ref{class2}), we find that the mass-radius relation for
classical white dwarf stars in $D$ dimensions is
\begin{equation}
M^{(D-2)/D}R^{4-D}={1\over 2}\biggl ({D\over 2 S_{D}^{2}}\biggr )^{2/D}{h^{2}\over mG(\mu H)^{(D+2)/ D}}\ \omega_{D/2}^{(D-2)/ D}.
\label{class9}
\end{equation}
For $2<D<4$, the mass $M$ decreases as the radius $R$ increases while for
$D<2$ and for $4<D<2(1+\sqrt{2})$ it increases with the radius (see
Fig. \ref{classique}).

For $D=4$ the mass is independent on the
radius and given in terms of fundamental constants by
\begin{equation}
M={\omega_{2}\over 2S_{4}^{2}}{h^{4}\over m^{2}G^{2}(\mu H)^{3}}\simeq 0.0143958...{h^{4}\over m^{2}G^{2}(\mu H)^{3}}.
\label{class10}
\end{equation}
We recall that the value of the gravitational constant $G$ depends on
the dimension of space so that we cannot give an explicit value to
this limiting mass. The central density is related to the radius by
\begin{equation}
\rho_{0}R^{4}={\xi_{1}^{4}\over 16\pi^{6}}{h^{4}\over m^{2}G^{2}(\mu H)^{3}}\simeq 0.105468... \ {h^{4}\over m^{2}G^{2}(\mu H)^{3}}.
\label{class10new}
\end{equation}
For $D=2$, the radius
is independent on mass and given in terms of fundamental constants by
\begin{equation}
R={\xi_{1}\over 2\sqrt{2}\pi}{h\over (Gm)^{1/2}\mu H}\simeq 0.270638...\ {h\over  (Gm)^{1/2}\mu H}.
\label{class11}
\end{equation}
Furthermore, for $D=2$, the Lane-Emden equation (\ref{class4}) with index
$n_{3/2}=1$ can be solved analytically, yielding
$\theta_{1}=J_{0}(\xi)$ and $\xi_{1}=\alpha_{0,1}$ where
$\alpha_{0,1}=2.404826...$ is the first zero of Bessel function
$J_{0}$. We obtain a density profile
\begin{equation}
\rho(r)=\rho_{0}J_{0}(\xi_{1}r/R),
\label{class12}
\end{equation}
where the central density is related to the total mass by
\begin{equation}
M=-{\rho_{0}\over 4\pi}{h^{2}\over Gm(\mu H)^{2}}\xi_{1}\theta'_{1}=0.0993492...{\rho_{0} h^{2}\over Gm(\mu H)^{2}}.
\label{class13}
\end{equation}
Finally, for $D=3$, using the mass-radius relation (\ref{class9}), we
find that the average density is related to the total mass by
$\overline{\rho}=2.162\ 10^{6}(M/M_{\odot})^2\ {\rm g}/{\rm cm}^{2}$ (for
$\mu=2.5$). Historically, this result was obtained by Chandrasekhar
(1931) \cite{chandra0} who first applied the theory of polytropic gas
spheres with index $n=3/2$ to classical white dwarf stars. It improves
an earlier result $\overline{\rho}=3.977\ 10^{6}(M/M_{\odot})^2\ {\rm g}/{\rm
cm}^{2}$ obtained by Stoner (1929) \cite{stoner0} on the basis of his
model of stars with uniform density (see Sec. \ref{sec_ana}).

\section{The ultra-relativistic  limit}
\label{sec_ur}

In the ultra-relativistic limit $x\gg 1$, we find that the equation of state takes the form
\begin{equation}
P=K_{2}\rho^{1+{1/ D}},
\label{ur1}
\end{equation}
with
\begin{equation}
 K_{2}={1\over D+1}\biggl ({D\over 2S_{D}}\biggr )^{1/D}{hc\over (\mu H)^{(D+1)/ D}}.
\label{ur2}
\end{equation}
Therefore, an ultra-relativistic white dwarf star is equivalent to a polytrope of index
\begin{equation}
n_{3}'={D}.
\label{ur3}
\end{equation}
This also directly results from the  Chandrasekhar equation. For $x\gg 1$, it reduces to
\begin{equation}
{1\over\xi^{D-1}}{d\over d\xi}\biggl (\xi^{D-1}{d\theta\over d\xi}\biggr )=-\theta^{D},
\label{ur4}
\end{equation}
\begin{equation}
\theta(0)=1, \qquad \theta'(0)=0,
\label{ur5}
\end{equation}
where we have set $\theta=\phi$ and $\xi=\eta$. This is the Lane-Emden
equation with index $n_{3}'=D$. 

\begin{figure}[htbp]
\centerline{
\includegraphics[width=8cm,angle=0]{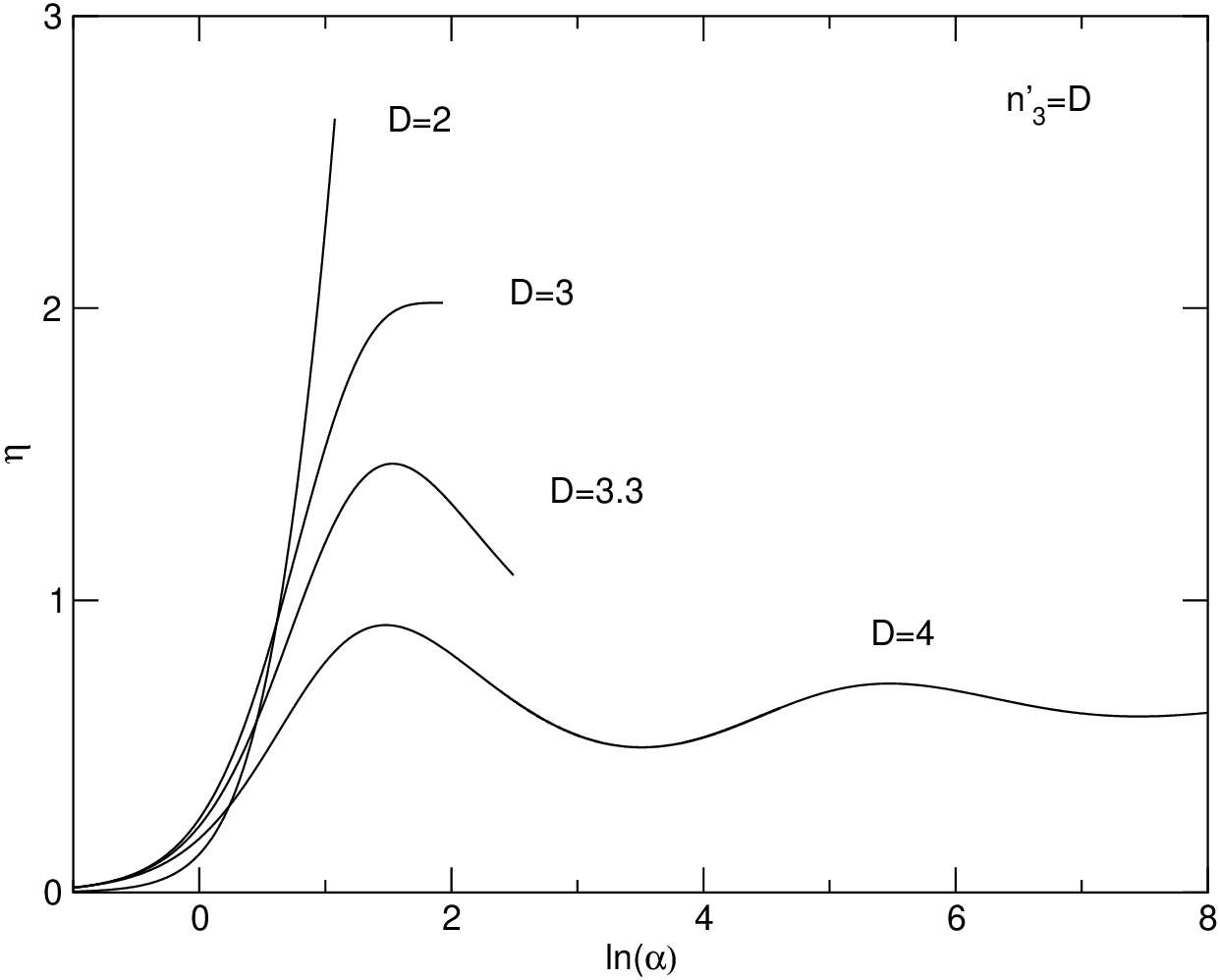}
} \caption[]{Relation between the mass (ordinate) and the central
density (abscissa) of box-confined polytropes with index $n_{3}'=D$ in
an appropriate system of coordinates (see \cite{lang,aaantonov} for
details). Complete polytropes correspond to the terminal point in the
series of equilibria. The series becomes dynamically unstable with
respect to the Euler-Poisson system (saddle point of the energy
functional) after the turning point of mass $\eta$ which appears for
$D=3$. Thus, ultra-relativistic white dwarf stars are stable for $D\le
3$ and unstable for $D>3$. }
\label{alphaR}
\end{figure}

\begin{figure}[htbp]
\centerline{
\includegraphics[width=8cm,angle=0]{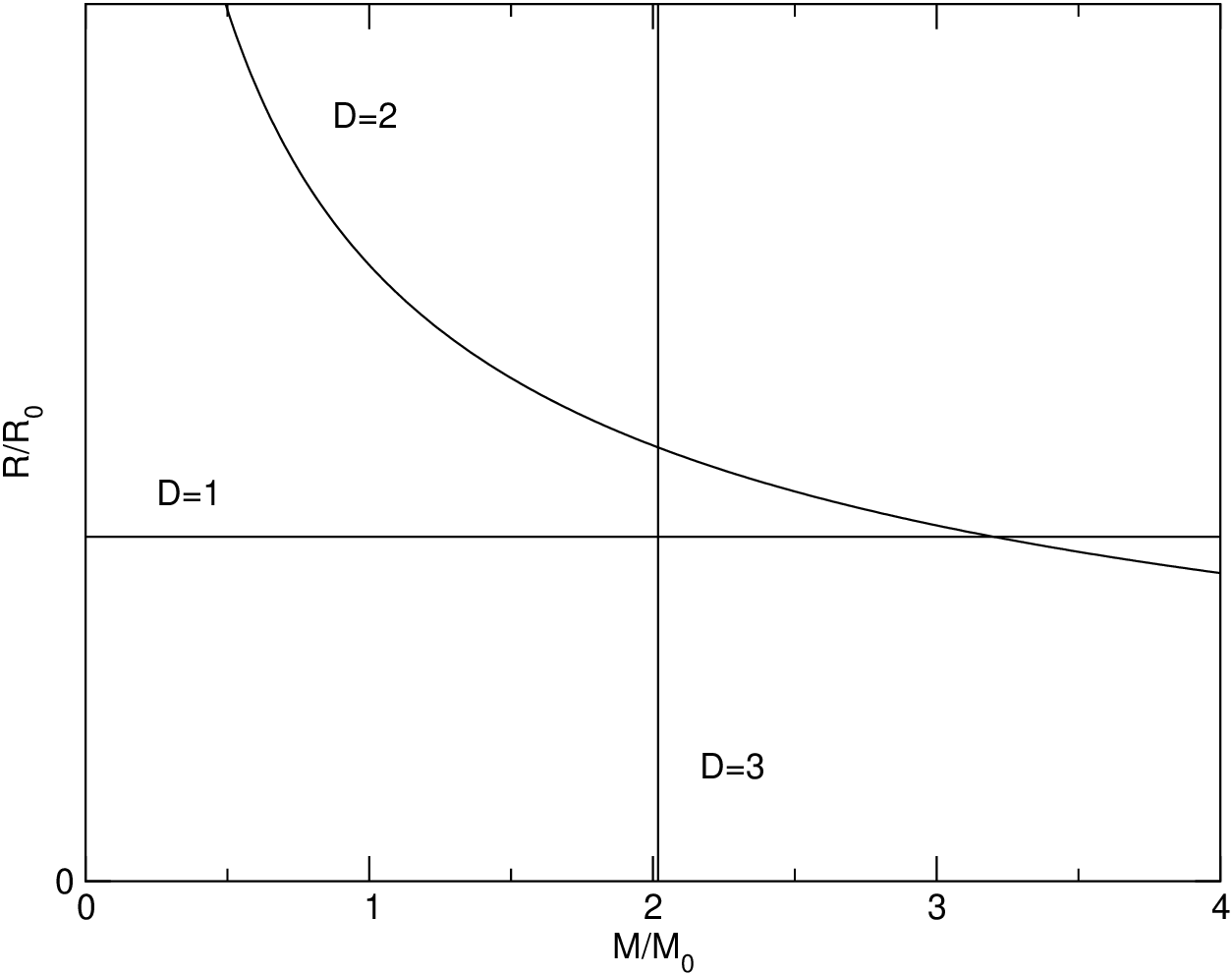}
} \caption[]{Mass-radius relation for ultra-relativistic white dwarf
stars in different dimensions of space. The radius is independent on
mass in $D=1$ and the mass is independent on radius in $D=3$
(Chandrasekhar's mass). } \label{relativiste}
\end{figure}

Using the results of \cite{lang} for $D>2$, we deduce that an
ultra-relativistic white dwarf star is self-confined if
$n_{3}'<n_{5}$, i.e. $D<(3+\sqrt{17})/2=3.5615528...$. In addition, it
is nonlinearly dynamically stable with respect to the Euler-Poisson
system if $n_{3}'<n_{3}$ and linearly unstable otherwise. Therefore,
ultra-relativistic white dwarf stars are stable for $D\le 3$ and
unstable for $D>3$ (see Fig. \ref{alphaR} and Appendix
\ref{sec_stab}). For $D\le 2$, ultra-relativistic white dwarf stars
are self-confined and stable. On the other hand, using the results of
\cite{lang}, the mass-radius relation for complete polytropes with
index $n_{3}'=D$ in $D$ dimensions is
\begin{equation}
M^{(D-1)/ D}R^{3-D}={K_{2}(D+1)\over GS_{D}^{1/D}}\omega_{D}^{(D-1)/ D},
\label{ur6}
\end{equation}
where we have defined
\begin{equation}
\omega_{D}=-\xi_{1}^{D+1\over D-1}\theta'_{D}(\xi_{1}).
\label{ur7}
\end{equation}
Using Eq. (\ref{ur2}), we find that the mass-radius relation for ultra-relativistic white dwarf stars in $D$ dimensions is
\begin{equation}
M^{(D-1)/ D}R^{3-D}=\biggl ({D\over 2S_{D}^{2}}\biggr )^{1/D}{hc\over G(\mu H)^{(D+1)/D}}\omega_{D}^{(D-1)/ D}.
\label{ur8}
\end{equation}
For $3<D<(3+\sqrt{17})/2$, the mass $M$ increases as the 
radius $R$ increases while for $1<D<3$ it decreases with the radius (see Fig.
\ref{relativiste}).

For $D=1$, the radius is independent on mass and
given in terms of fundamental constants by
\begin{equation}
R={\xi_{1}\over \sqrt{2}S_{1}}\biggl ({hc\over G}\biggr )^{1/2}
{1\over\mu H}=0.555360...\biggl ({hc\over G}\biggr )^{1/2}{1\over\mu
H}. \label{ur9}
\end{equation}
Furthermore, for $D=1$, the Lane-Emden equation (\ref{ur4}) with index
$n_{3}'=1$ can be solved analytically, yielding
$\theta_{1}=\cos(\xi)$ and $\xi_{1}=\pi/2$. We obtain a density
profile
\begin{equation}
\rho(r)=\rho_{0}\cos(\xi_{1}r/R), \label{ur10}
\end{equation}
where the central density is related to the total mass by
\begin{equation}
M={\rho_{0}\over \sqrt{2}}\left ({h c\over G}\right )^{1/2}{1\over
\mu H}. \label{ur11}
\end{equation}
For $D=3$, the mass is independent on radius and given in terms of
fundamental constants by
\begin{equation}
M=\biggl ({3\over 32\pi^{2}}\biggr )^{1/2}\omega_{3}\biggl ({hc\over G}\biggr )^{3/2}{1\over (\mu H)^{2}}.
\label{ur12}
\end{equation}
This is the Chandrasekhar mass
\begin{equation}
M=0.196701...\biggl ({hc\over G}\biggr )^{3/2}{1\over (\mu H)^{2}}\simeq 5.76 M_{\odot}/\mu^{2}.
\label{ur13}
\end{equation}
Coming back to Eq. (\ref{c7}), we can show  that for this limiting value,
the radius $R$ of the configuration tends to zero (see
Sec. \ref{sec_gen}). Historically, the existence of a maximum mass for
relativistic white dwarf stars was first published by Anderson (1929)
\cite{anderson} who considered a relativistic extension of the model
of Stoner (1929) \cite{stoner0} for classical white dwarf stars. He
obtained a limiting mass $M=1.37\ 10^{33}\ {\rm g}$ (for
$\mu=2.5$). The relativistic treatment of Anderson was criticized and
corrected by Stoner (1930) \cite{stoner} who obtained a value of the
limiting mass $M=2.19\ 10^{33}\ {\rm g}$. The uniform density model of
Stoner was in turn criticized and corrected by Chandrasekhar (1931)
\cite{chandra1} who applied the theory of polytropic gas spheres with
index $n=3$ to relativistic white dwarf stars and obtained the value
(\ref{ur13}) of the limiting mass $M=1.822\ 10^{33}\ {\rm g}$. It
seems that these historical details are not well-known because the
references to the works of Anderson and Stoner progressively
disappeared from the literature.

\section{The general case}
\label{sec_gen}

Collecting together the results of Sec. \ref{sec_c}, the mass-radius relation
for relativistic white dwarf stars in $D$ dimensions can be written
in the general case under the parametric form
\begin{equation}
{M\over M_{0}}=y_{0}^{{D}(3-D)/2}\Omega(y_{0}),\qquad {R\over R_{0}}={1\over y_{0}^{{(D-1)/2}}}\eta_{1},
\label{gen1}
\end{equation}
where we have defined
\begin{equation}
M_{0}=S_{D}\biggl ({8A_{2}\over S_{D}G}\biggr )^{D/2}{1\over B^{D-1}},\quad R_{0}=\biggl ({8A_{2}\over S_{D}G}\biggr )^{1/2}{1\over B},
\label{gen2}
\end{equation}
and
\begin{equation}
\Omega(y_{0})=-\biggl (\eta^{D-1}{d\phi\over d\eta}\biggr )_{\eta=\eta_{1}}.
\label{gen3}
\end{equation}
The mass $M_{0}$ and the radius $R_{0}$ can be expressed in terms of fundamental constants as
\begin{equation}
R_{0}=\biggl ({D\over 2S_{D}^{2}}\biggr )^{1/2}{h^{D/2}\over G^{1/2}m^{(D-1)/2}c^{(D-2)/2}\mu H},
\label{gen4}
\end{equation}
\begin{equation}
M_{0}=\left ({D\over 2S_{D}^{2}}\right )^{(D-2)/2}{h^{D(D-2)/2}c^{(4-D)D/2} \over m^{(D-3)D/2}  G^{D/2}}{1\over (\mu H)^{D-1}}.
\label{gen5}
\end{equation}
We can now obtain the mass-radius curve $M-R$ by the following
procedure. We fix a value of the parameter $y_0$ and solve the
differential equation (\ref{c7}) with initial condition (\ref{c8}) until the point
$\eta=\eta_{1}$, determined by Eq. (\ref{c9}), at which the density
vanishes. The radius and the mass of the corresponding configuration
are then given by Eq. (\ref{gen1}). By varying $y_{0}$, we can
obtain the full curve $R(y_{0})-M(y_{0})$ parameterized by the value
of the central density $\rho_0$ given by Eq. (\ref{c11}). To solve the
differential equation (\ref{c7}), we need the behavior of $\phi$ at the
origin. Expanding $\phi(\eta)$ in Taylor series and substituting
this expansion in Eq. (\ref{c7}) we obtain for $\eta\rightarrow 0$:
\begin{equation}
\phi=1-{q^{D}\over 2D}\eta^{2}+{1\over 8(D+2)}q^{2(D-1)}\eta^{4}+...
\label{gen6}
\end{equation}
where
\begin{equation}
q^{2}=1-{1\over y_{0}^{2}}.
\label{gen7}
\end{equation}
We note in particular that $\phi''(0)=-{q^{D}/ D}$.

\begin{figure}[htbp]
\centerline{
\includegraphics[width=8cm,angle=0]{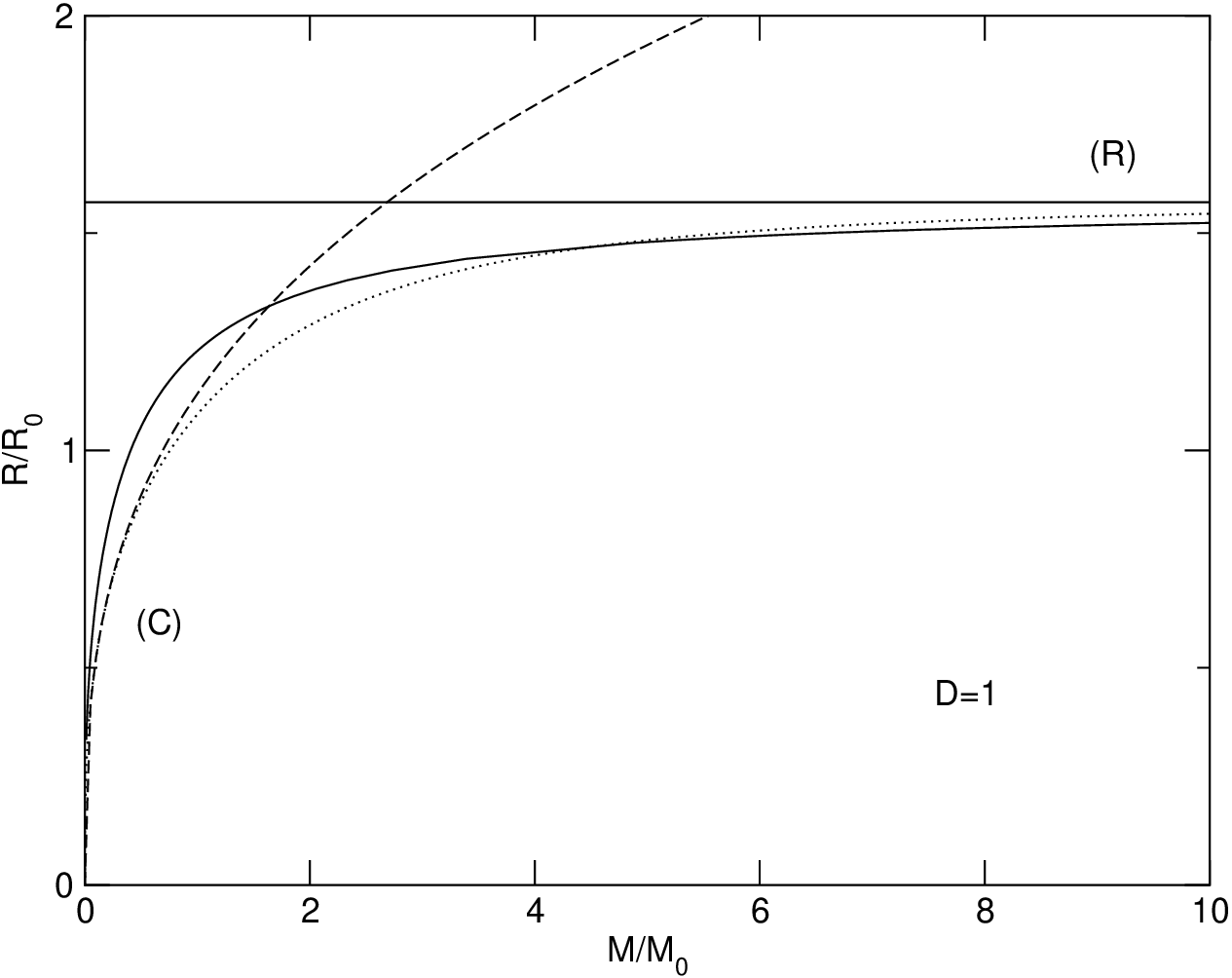}
} \caption[]{Mass-radius relation in $D=1$. The radius increases
with the mass. The configurations are always stable and there exists
a maximum radius achieved in the ultra-relativistic limit for
$M\rightarrow +\infty$. } \label{chandra1}
\end{figure}

\begin{figure}[htbp]
\centerline{
\includegraphics[width=8cm,angle=0]{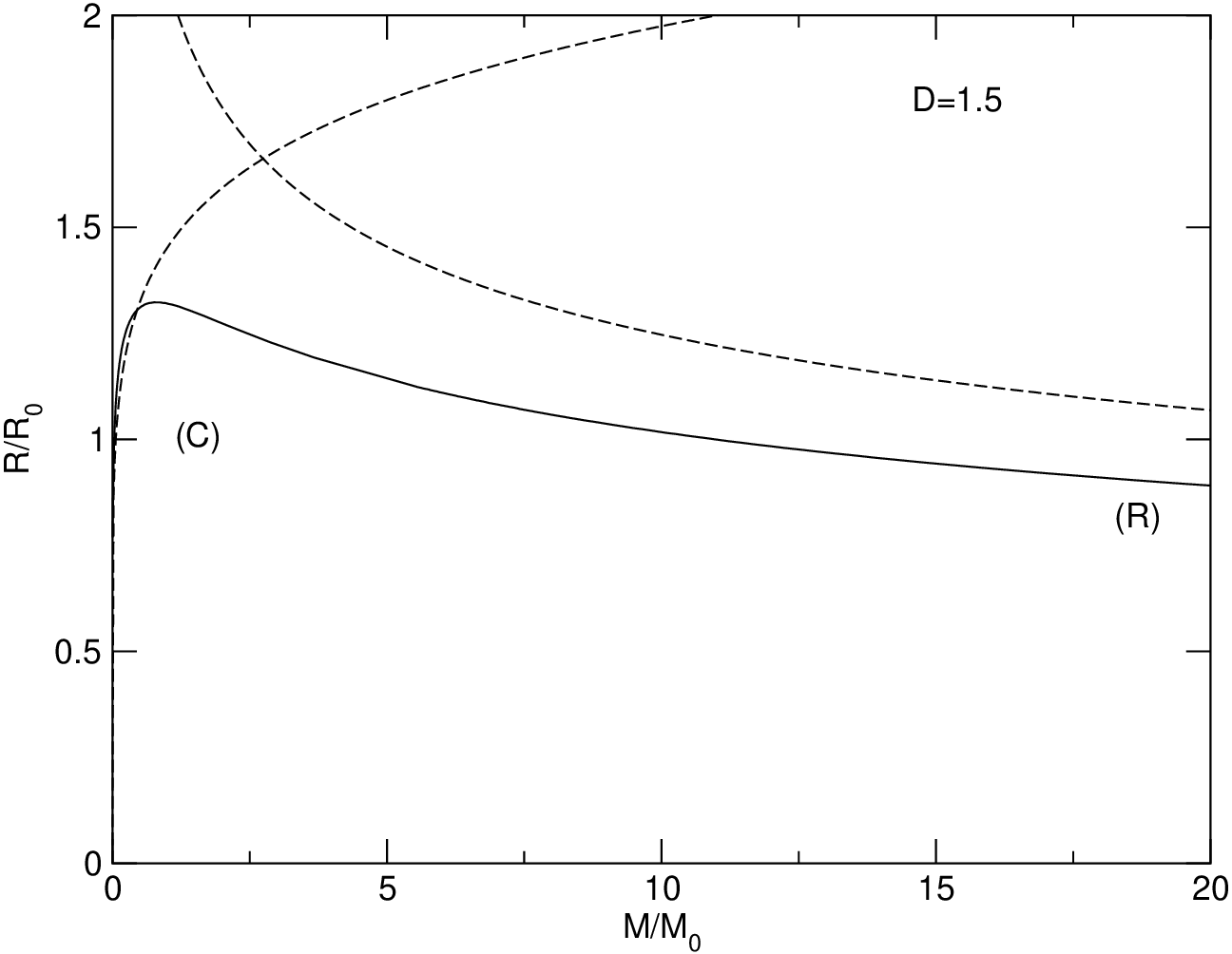}
} \caption[]{Mass-radius relation in $D=1.5$. The configurations are
always stable and there exists a maximum radius for partially
relativistic distributions. } \label{chandra1p5}
\end{figure}

\begin{figure}[htbp]
\centerline{
\includegraphics[width=8cm,angle=0]{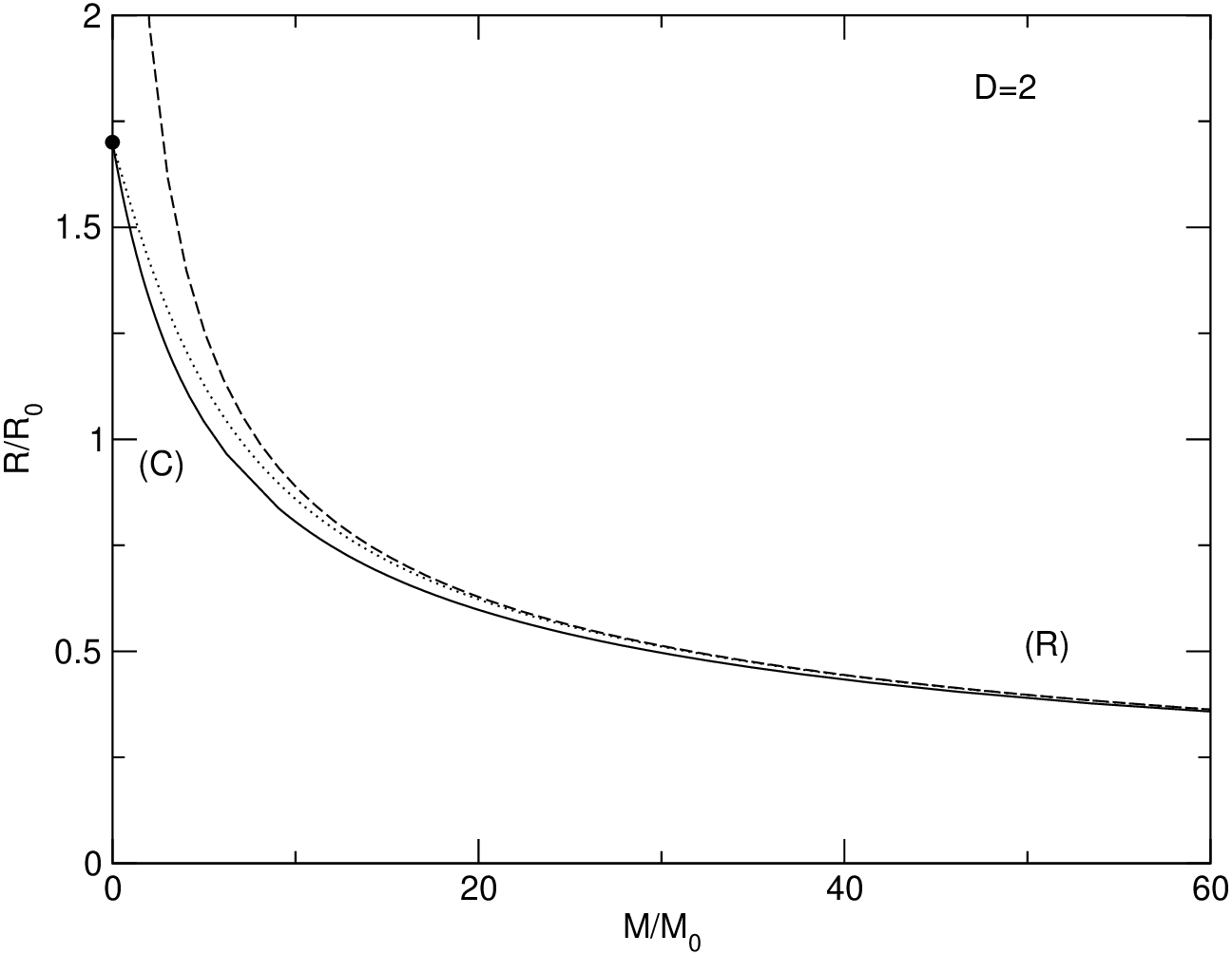}
} \caption[]{Mass-radius relation in $D=2$. The radius decreases
with the mass. The configurations are always stable and there exists
a maximum radius achieved in the classical limit for $M\rightarrow
0$. } \label{chandra2}
\end{figure}

\begin{figure}[htbp]
\centerline{
\includegraphics[width=8cm,angle=0]{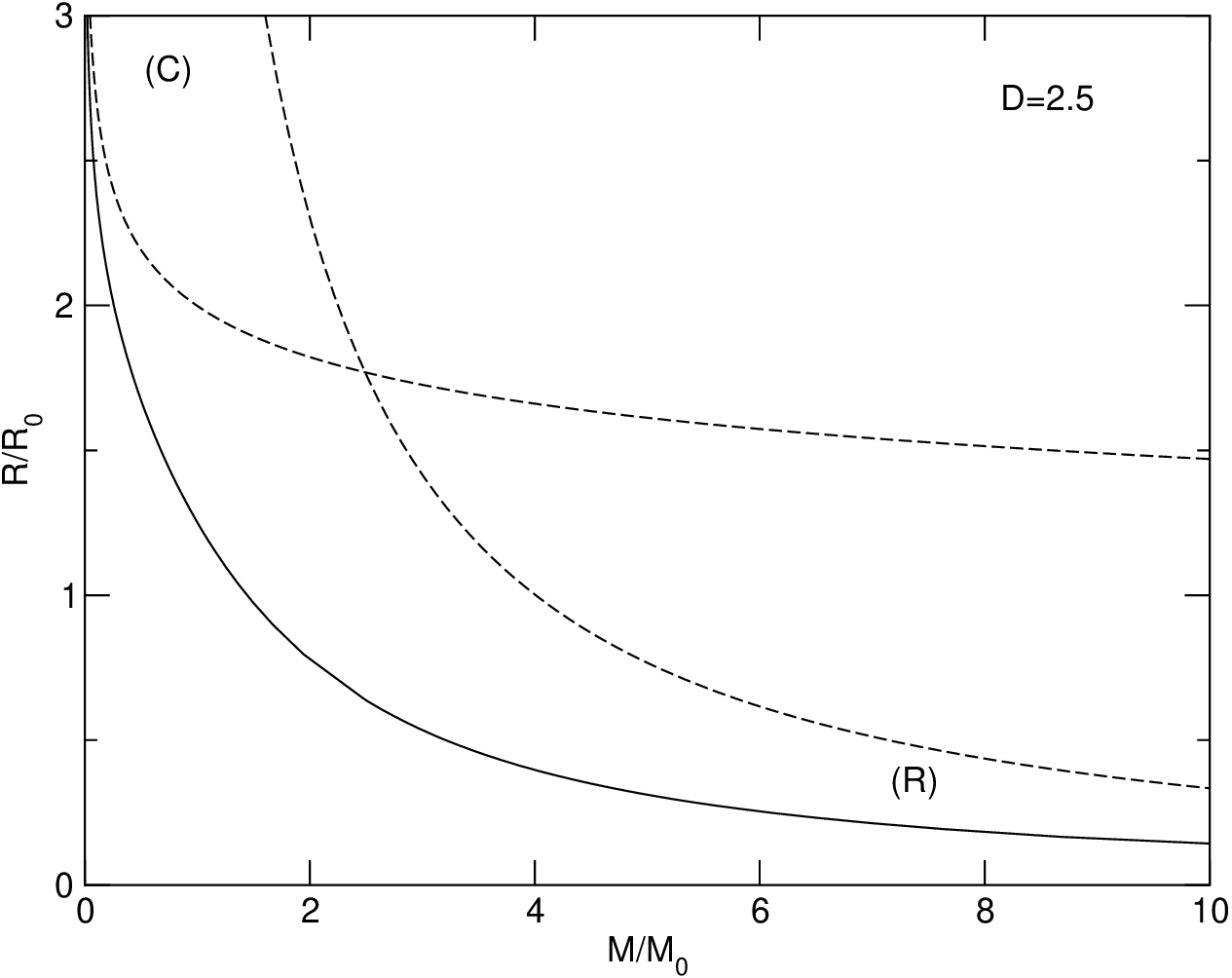}
} \caption[]{Mass-radius relation in $D=2.5$. The radius decreases
with the mass. The configurations are always stable.  }
\label{chandra2p5}
\end{figure}

\begin{figure}[htbp]
\centerline{
\includegraphics[width=8cm,angle=0]{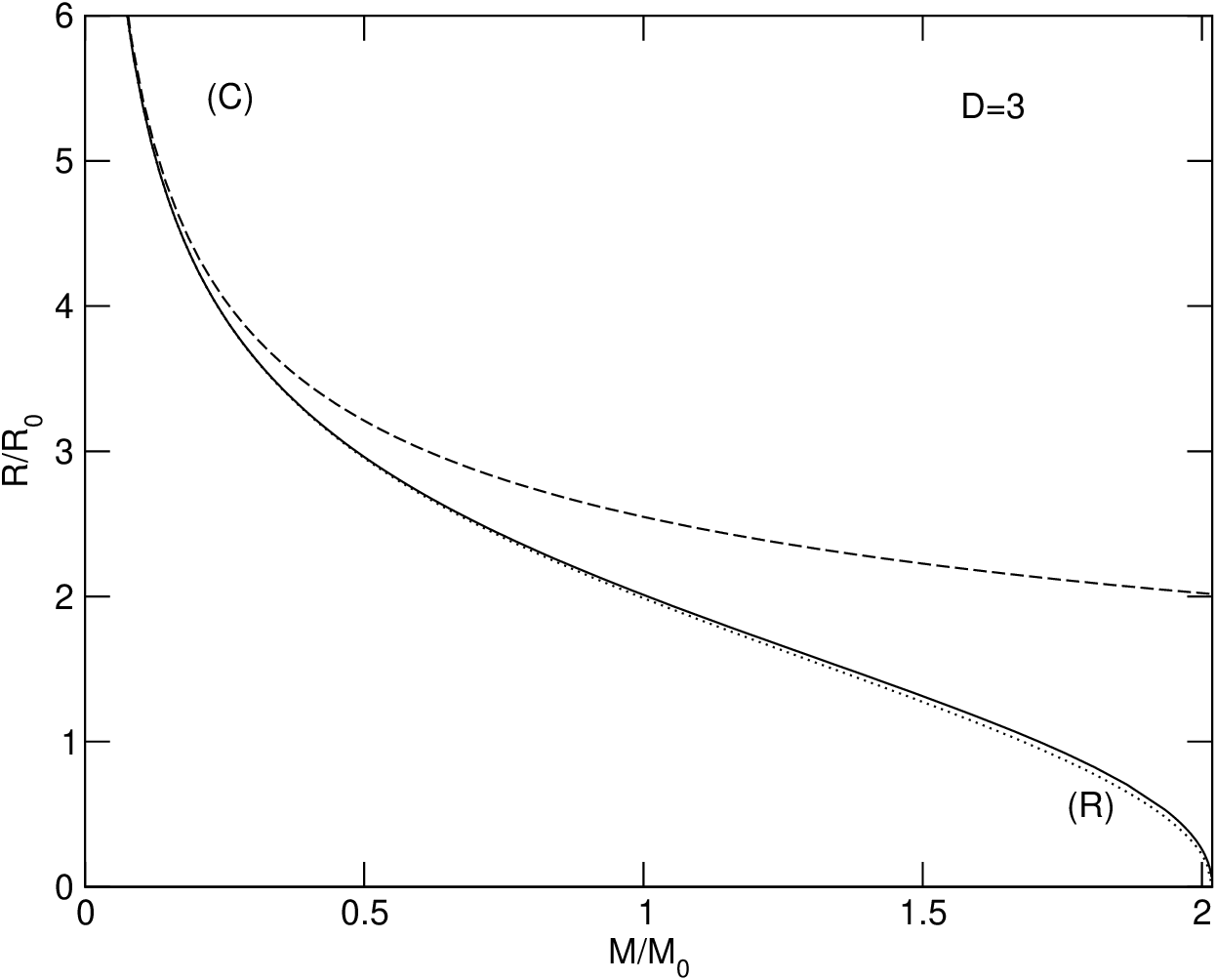}
} \caption[]{Mass-radius relation in $D=3$. The radius decreases
with the mass. The configurations are always stable and there exists
a limiting mass (Chandrasekhar's mass) achieved in the
ultra-relativistic limit for $R=0$. For $M>M_{Chandra}$, quantum
mechanics cannot arrest gravitational collapse. } \label{chandra3}
\end{figure}

\begin{figure}[htbp]
\centerline{
\includegraphics[width=8cm,angle=0]{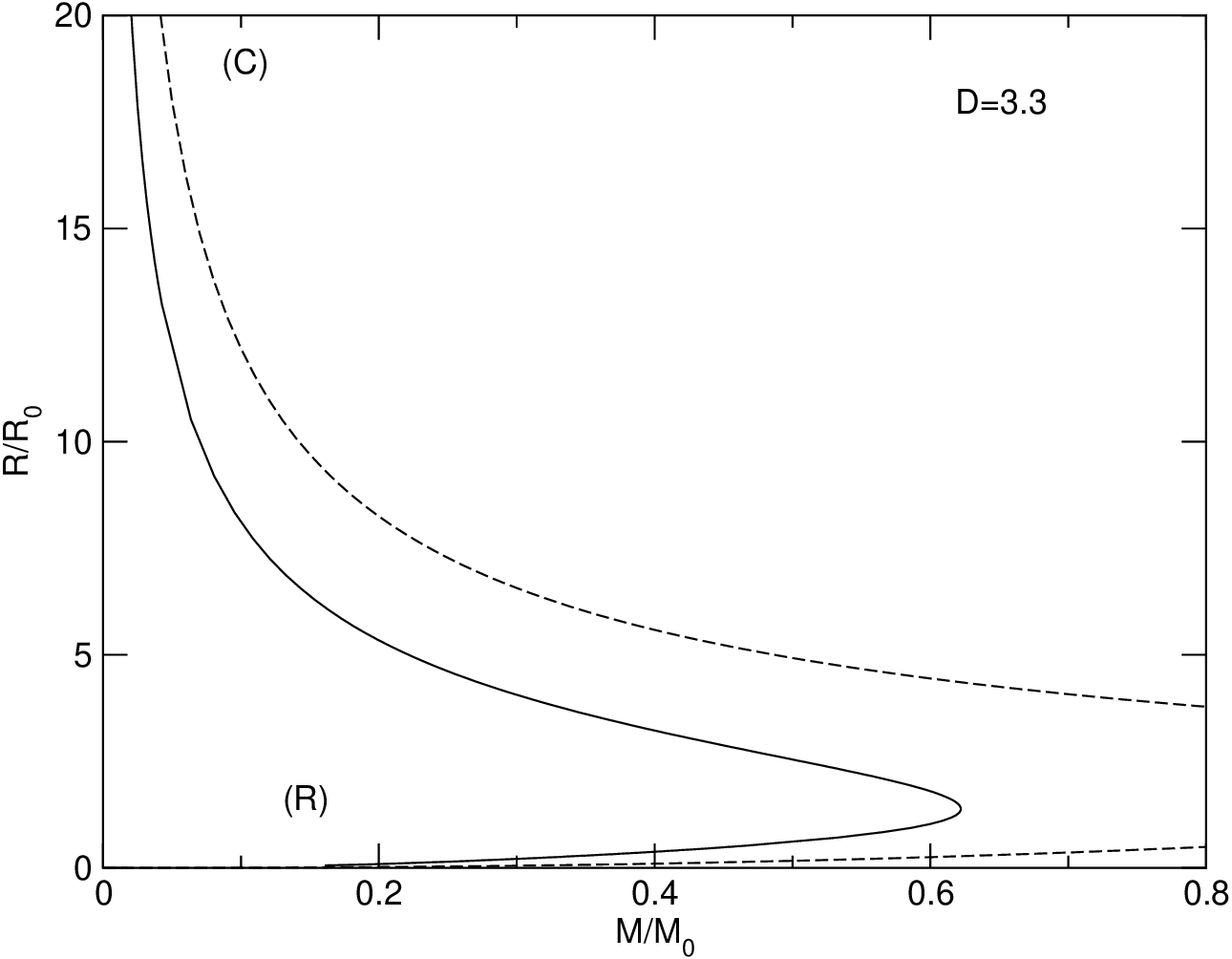}
} \caption[]{ Mass-radius relation in $D=3.3$. There exists a
limiting mass for partially relativistic distributions. Classical
configurations are stable and the series of relativistic equilibria
becomes unstable after the turning point of mass. Thus, highly
relativistic configurations in $D>3$ cannot be in hydrostatic
equilibrium.} \label{chandra3p3}
\end{figure}

\begin{figure}[htbp]
\centerline{
\includegraphics[width=8cm,angle=0]{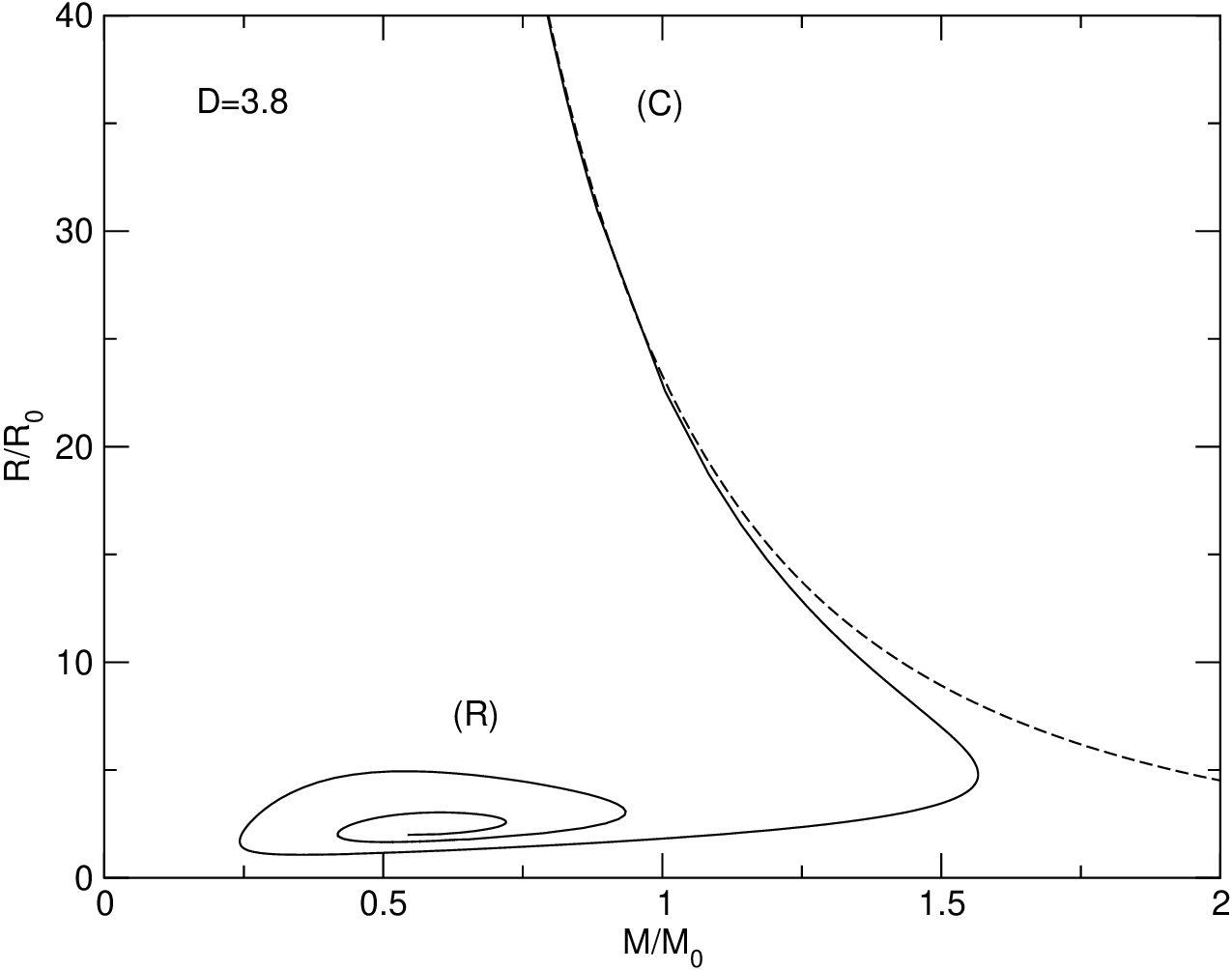}
} \caption[]{Mass-radius relation in $D=3.8$. For $D>(3+\sqrt{17})/2$ the ultra-relativistic configurations (equivalent to
polytropes of index $n'_3=D$) are not self-confined anymore and a
spiral develops in the mass-radius relation. This is somehow similar
to the classical spiral occurring in the $(E,\beta)$ plane in the
thermodynamics of isothermal self-gravitating systems 
\cite{lbw,katz,paddy,iso}  and
to the spiral occurring in the $(M,R)$ plane in the general
relativistic treatment of neutron stars \cite{meltzer,relativity}.  The series of equilibria becomes unstable at
the first turning point of mass and new modes of instability occur at
the secondary turning points \cite{shapiro}.} \label{chandra3p8}
\end{figure}

\begin{figure}[htbp]
\centerline{
\includegraphics[width=8cm,angle=0]{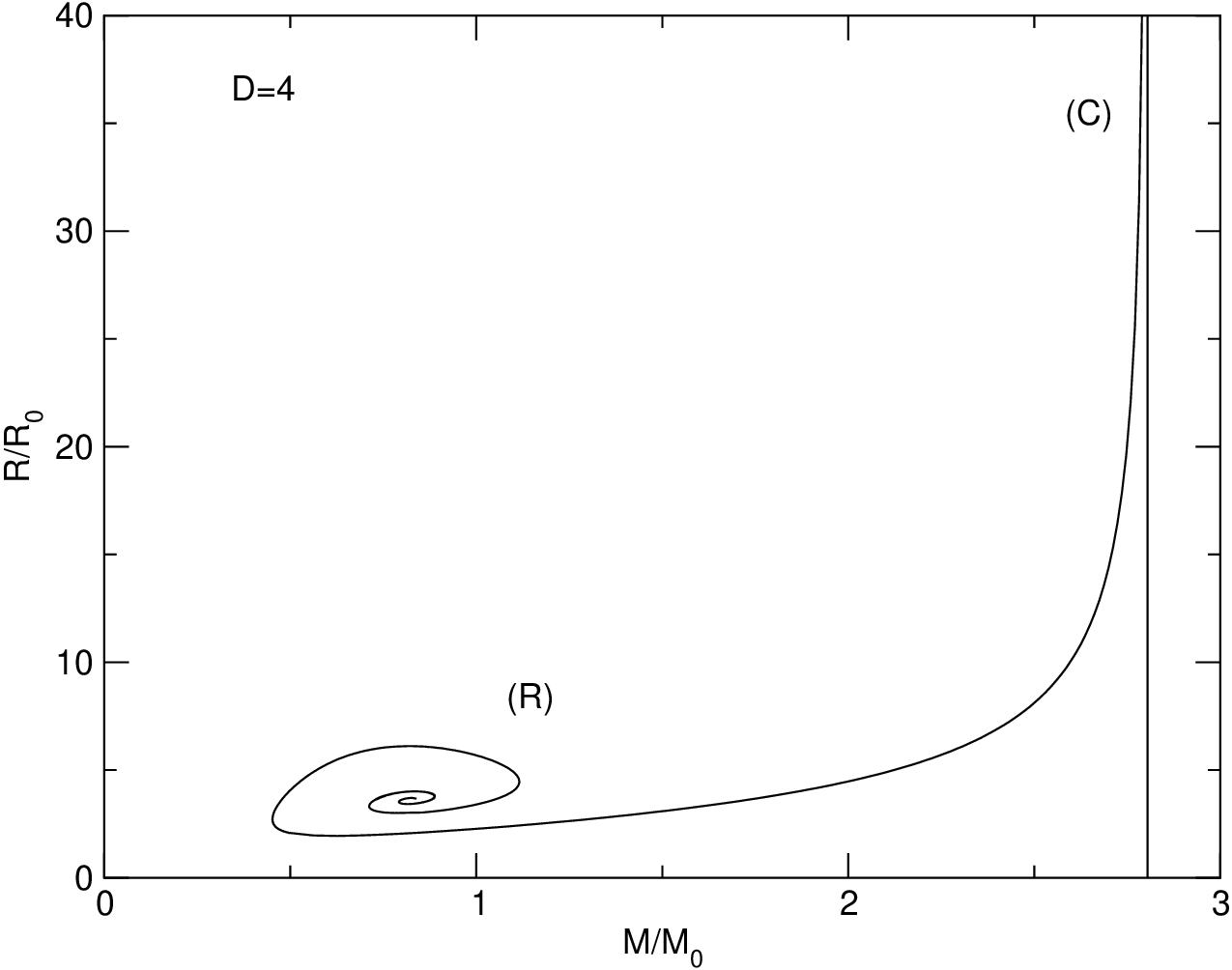}
} \caption[]{Mass-radius relation in $D=4$. There exists a limiting
mass achieved in the classical limit for $R\rightarrow +\infty$. In
fact, for $D\ge 4$, all the configurations (classical and
relativistic) are unstable. Quantum mechanics cannot balance
gravitational attraction even in the classical limit. }
\label{chandra4}
\end{figure}

Finally, we give the asymptotic expressions of the mass-radius relation.
In the classical limit, using Eq. (\ref{class9}), we obtain
\begin{equation}
\left ({M\over M_{0}}\right )^{(D-2)/D}\left ({R\over R_{0}}\right )^{4-D}={1\over 2}\omega_{D/2}^{(D-2)/D}.
\label{gen8}
\end{equation}
In the ultra-relativistic limit, using Eq. (\ref{ur8}), we get
\begin{equation}
\left ({M\over M_{0}}\right )^{(D-1)/D}\left ({R\over R_{0}}\right )^{3-D}=\omega_{D}^{(D-1)/D}.
\label{gen9}
\end{equation}
In Figs. \ref{chandra1}-\ref{chandra4}, we plot the mass-radius
relation of relativistic white dwarf stars (full line) for different
dimensions of space. The asymptotic relations (\ref{gen8}) and
(\ref{gen9}) valid in the classical (C) and ultra-relativistic (R)
limits are also shown for comparison (dashed line) together with the
analytical approximation (dotted line) derived in Sec. \ref{sec_ana}.

\section{Analytical approximation of the mass-radius relation}
\label{sec_ana}

We present here an analytical approximation of the mass-radius
relation in various dimensions of space based on the treatment by
Nauenberg \cite{nauenberg} in $D=3$. This treatment amounts to
considering that the density is uniform in the star and the
mass-radius relation is obtained by minimizing the energy functional
(see Appendix \ref{sec_ef}) with respect to the radius (or to the
density) at fixed mass. As mentioned in the Introduction, this is
similar to the simplified model of relativistic white dwarf stars made
by Stoner \cite{stoner} before Chandrasekhar's treatment
\cite{chandra1}. Following Nauenberg
\cite{nauenberg}, we approximate the kinetic energy $K$ by the form
\begin{equation}
K=N mc^{2}\biggl\lbrace \biggl (1+{p^{2}\over m^{2}c^{2}}\biggr )^{1/2}-1\biggr\rbrace,
\label{ana1}
\end{equation}
where $N$ is the number of electrons and $p$ is an average over the
star of the momentum of the electrons \footnote{Note that Stoner
\cite{stoner} uses the exact expression (\ref{es2}) of the 
kinetic energy in his treatment. This leads to more complicated
expressions with, however, qualitatively similar conclusions. Since
the models are based on the (unrealistic) approximation that the
stellar density is homogeneous, the results cannot be expected to be
more than qualitatively correct.  Therefore, the approximation
(\ref{ana1}) made by Nauenberg for the kinetic energy is sufficient
for the purposes of this simplified approach.}. We assume that it is
determined by an appropriate average value of the density by the
relation
\begin{equation}
\rho={2S_{D}\over Dh^{D}}\mu H p^{D}, \label{ana2}
\end{equation}
based on the Pauli exclusion principle. Now, for the density, we
write
\begin{equation}
\rho=\zeta {M \over S_{D}R^{D}}, \label{ana3}
\end{equation}
where $\zeta$ is a dimensionless parameter. We also write the potential energy in the form
\begin{equation}
W=-{\nu\over D-2}{GM^{2}\over R^{D-2}},
\label{ana4}
\end{equation}
where $\nu$ is another dimensionless parameter. By writing Eq. (\ref{ana4}),
we have assumed that $D\neq 2$ but we shall see that the following
results pass to the limit for $D\rightarrow 2$. We introduce two
dimensionless variables $n$ and $r$ and two fixed constants
$M_{*}=N_* \mu H$ and $R_{*}$ such that
\begin{equation}
M=n M_{*}, \qquad R=rR_{*}.
\label{ana5}
\end{equation}
We determine $M_{*}$ and $R_{*}$ by the requirement that the
relativity parameter $x=p/mc$ have the form
\begin{equation}
x={n^{1/D}\over r},
\label{ana6}
\end{equation}
and that the potential energy can be written
\begin{equation}
W=-{1\over D-2}mc^{2}{N_{*}}{n^{2}\over r^{D-2}}.
\label{ana7}
\end{equation}
This yields
\begin{equation}
R_{*}=\biggl ({D\zeta\over 2\nu S_{D}^{2}}\biggr )^{1/2}{h^{D/2}\over G^{1/2}m^{(D-1)/2}c^{(D-2)/2}\mu H},
\label{ana8}
\end{equation}
\begin{equation}
M_{*}={1\over \nu^{D/2}}\biggl ({D\zeta\over 2S_{D}^{2}}\biggr )^{(D-2)/2}
{h^{D(D-2)/ 2}c^{D(4-D)/ 2}\over m^{D(D-3)/ 2}G^{D/2}}{1\over (\mu H)^{D-1}}.
\label{ana9}
\end{equation}
Comparing with Eqs. (\ref{gen4}) and  (\ref{gen5}), we find that
\begin{equation}
M_{*}={1\over \nu^{D/2}}\zeta^{(D-2)/ 2}M_{0}, \quad R_{*}=\biggl ({\zeta\over \nu}\biggr )^{1/2}R_{0}.
\label{ana10}
\end{equation}
Now, the energy $E=K+W$ of the star can be written
\begin{equation}
E=N_{*}mc^{2}n\biggl\lbrace (1+x^{2})^{1/2}-1-n^{2/D}{x^{D-2}\over D-2}\biggr\rbrace.
\label{ana11}
\end{equation}
In the classical limit $x\ll 1$, we get
\begin{equation}
E_{class}=N_{*}mc^{2}\left\lbrace {n^{1+2/D}\over 2 r^{2}}-{n^{2}\over (D-2)r^{D-2}}\right\rbrace,
\label{ana11class}
\end{equation}
and in the ultra-relativistic limit $x\gg 1$, we obtain
\begin{equation}
E_{relat}=N_{*}mc^{2}\left\lbrace {n^{1+1/D}\over  r}-{n^{2}\over (D-2)r^{D-2}}\right\rbrace.
\label{ana11relat}
\end{equation}
We shall consider the energy as a function of the radius $R$, with
the  mass $M$ fixed. Thus, the mass-radius relation will be obtained
by minimizing the energy $E$ versus $x$ with fixed ${n}$. Writing
$\partial E/\partial x=0$, we obtain the equations
\begin{equation}
{n}={x^{{D}(4-D)/2}\over (1+x^{2})^{D/4}},
\label{ana12}
\end{equation}
\begin{equation}
x={{n}^{1/D}\over r},
\label{ana13}
\end{equation}
defining the mass-radius relation in parametric form in the framework
of the simplified model. Note that $x^D$ represents the value of the
density in units of $M_*/R_{*}^D$, i.e. $\rho=x^D
M_*/R_{*}^D$. Therefore, Eq. (\ref{ana12}) can be viewed as the
relation between the mass and the density. Eliminating the relativity
parameter $x$ (or density) between Eqs. (\ref{ana12}) and
(\ref{ana13}), we explicitly obtain
\begin{equation}
{n}={r^{3}\over \sqrt{1-r^{4}}} \qquad (D=1)
\label{ana14}
\end{equation}
\begin{equation}
{n}={1-r^{4}\over r^{2}} \qquad (D=2)
\label{ana15}
\end{equation}
\begin{equation}
r={(1-{n}^{4/3})^{1/2}\over {n}^{1/3}} \qquad (D=3)
\label{ana16}
\end{equation}
\begin{equation}
r={n^{3/4}\over \sqrt{1-n}} \qquad (D=4)
\label{ana17}
\end{equation}
In $D=1$, there exists a maximum radius $r=1$ achieved for $n\rightarrow +\infty$, and we have the scaling 
\begin{equation}
n\sim \frac{1}{2}(1-r)^{-1/2}.
\end{equation}
In $D=2$, there exists a maximum radius $r=1$ achieved for $n\rightarrow 0$, and we have the scaling 
\begin{equation}
n\sim 4(1-r).
\end{equation}
In $D=3$, there exists a maximum mass $n=1$ (Chandrasekhar's mass) achieved for $r\rightarrow 0$, and we have the scaling 
\begin{equation}
r\sim \frac{2}{\sqrt{3}}(1-n)^{1/2}.
\end{equation}
In $D=4$, there exists a maximum mass $n=1$  achieved for $r\rightarrow +\infty$, and we have the scaling 
\begin{equation}
r\sim (1-n)^{-1/2}.
\end{equation}
In order to compare Eqs. (\ref{ana14})-(\ref{ana17}) with the exact
results, we need to estimate the values of the constants $\zeta$ and
$\nu$. This can be done by considering the limiting forms of
Eqs. (\ref{ana12}) and (\ref{ana13}). In the classical limit $x\ll 1$,
we obtain
\begin{equation}
n^{(D-2)/ D}r^{4-D}=1,
\label{ana18}
\end{equation}
and in the ultra-relativistic limit $x\gg 1$, we get
\begin{equation}
n^{(D-1)/D}r^{3-D}=1.
\label{ana19}
\end{equation}
Comparing with Eqs. (\ref{gen8}) and (\ref{gen9}), we find that
\begin{equation}
\zeta=\left ({1\over 2}\right )^{D} {\omega_{D/2}^{D-2}\over \omega_{D}^{D-1}},\qquad \nu={1\over 2} {\omega_{D/2}^{(D-2)/D}\over \omega_{D}^{2(D-1)/D}},
\label{ana20}
\end{equation}
where the quantities $\omega_{D}$ and $\omega_{D/2}$ can be deduced
from the numerical study of the Lane-Emden equation. The analytical
approximations of the mass-radius relation (\ref{ana14})-(\ref{ana17})
are plotted in dotted lines in Figs. \ref{chandra1}, \ref{chandra2},
\ref{chandra3}, and they give a fair agreement with the exact results
(full line).  Of course, they cannot reproduce the spiral in $D=4$
which requires the resolution of the full differential equation (\ref{c7}).

Let us address the stability of the configurations within this
simplified analytical model. If we view a white dwarf star as a gas of
electrons at statistical equilibrium at temperature $T$, stable
configurations are those that minimize the free energy $F=E-TS$ at
fixed mass, where $E$ is the total energy (kinetic $+$ potential) and
$S$ is the Fermi-Dirac entropy (see Appendix \ref{sec_ef}). For a
completely degenerate gas at $T=0$, stable configurations are those
that minimize the energy $E$ at fixed mass. As we have seen
previously, the first order variations $E'(x)=0$ determine the
mass-radius relation.  Then, the configurations is stable if the
energy is minimum, i.e. $E''(x)>0$.  Now, we have
\begin{equation}
E''={N_{*}mc^{2}{n}\over (1+x^{2})^{3/2}}\lbrack 4-D-(D-3)x^{2}\rbrack.
\label{ana21}
\end{equation}
For $D\le 3$, all the solutions are stable while for $D\ge 4$, all
the solutions are unstable. For $3<D< 4$, the solutions with
$x<x_{c}$ are stable and the solutions with  $x>x_{c}$ are unstable
where
\begin{equation}
x_{c}=\biggl ({4-D\over D-3}\biggr )^{1/2}. \label{ana22}
\end{equation}
We now show that the onset of instability precisely corresponds to
the  turning point of mass, i.e. to the point where the mass is
maximum in the series of equilibria $M(R)$. In terms of reduced
variables, this corresponds to $dn/dr=0$, or equivalently $n'(x)=0$.
Taking the logarithmic derivative of Eq. (\ref{ana12}), we have
\begin{equation}
{dn\over n}={D\over 2}(4-D){dx\over x}-{D\over 2}{x\over 1+x^{2}}dx,
\label{ana23}
\end{equation}
so that the condition $n'(x)=0$ yields $x=x_{c}$. Thus, instability
sets in precisely at the maximum  mass as could have  been directly
inferred from the turning point criterion \cite{shapiro}.

It is also useful to discuss the classical and the ultra-relativistic
limits specifically. In the classical  limit, using Eq. (\ref{ana11class}), we get
\begin{equation}
E_{class}''=N_{*}mc^{2}n(4-D),
\label{ana23class}
\end{equation}
so that a classical white dwarf star is stable for $D<4$ and unstable
for $D\ge 4$. In the ultra-relativistic limit, using
Eq. (\ref{ana11relat}), we get
\begin{equation}
E_{relat}''={N_{*}mc^{2}n\over x}(3-D),
\label{ana23relat}
\end{equation}
so that an ultra-relativistic white dwarf star is stable for $D\le 3$ and
unstable for $D>3$. This returns the results obtained in Secs. \ref{sec_class} and \ref{sec_ur}.

Finally, considering the function $E(R)$ given by Eq. (\ref{ana11})
and assuming that the system evolves so as to minimize its energy
(this requires some source of dissipation), we have the following
results \footnote{We do not plot the curves $E(R)$ because they are
elementary, but drawing them is helpful to visualize the results described in
the text.}: for $D<3$, there exists a stable equilibrium state (global
minimum of $E$) with radius $R$ for all mass $M$. For $D=3$, there
exists a maximum mass $M_{c}$ so that: (i) for $M<M_{c}$, there exists
a stable equilibrium state (global minimum of $E$) with radius $R>0$
(ii) for $M=M_{c}$, the system collapses to a point ($R=0$) but its
energy remains finite (lower bound) (iii) for $M>M_{c}$, the system
collapses to a point $R\rightarrow 0$ and $E\rightarrow -\infty$. For
$3<D<4$, there exists a maximum mass $M_{c}$ so that: (i) for
$M<M_{c}$, there exists a metastable equilibrium state (local minimum
of $E$) with $x<x_{c}$ and an unstable equilibrium state (global
maximum of $E$) with $x>x_{c}$. The system can either reach the
metastable state or collapse to a point ($R\rightarrow 0$,
$E\rightarrow -\infty$); the choice probably depends on a notion of
basin of attraction (ii) for $M\ge M_{c}$, the system collapses to a
point ($R\rightarrow 0$, $E\rightarrow -\infty$). For $D= 4$, there
exists a critical mass $M_{c}$ so that: (i) for $M<M_{c}$, there
exists an unstable equilibrium state (global maximum of $E$) so the
system either collapses ($R\rightarrow 0$ and $E\rightarrow -\infty$)
or evaporates ($R\rightarrow +\infty$ and $E\rightarrow 0$) (ii) for
$M>M_{c}$, the system collapses to a point ($R\rightarrow 0$ and
$E\rightarrow -\infty$). For $D>4$, there exists an unstable
equilibrium state (global maximum of $E$) for all mass $M$ so the
system either collapses ($R\rightarrow 0$ and $E\rightarrow -\infty$)
or evaporates ($R\rightarrow +\infty$ and $E\rightarrow 0$).

\section{Conclusion} \label{sec_model}

{\it ``Why is our universe three dimensional? Does the dimension $D=3$
play a special role among other space dimensions?''}

Several scientists have examined the role played by the dimension of
space in determining the form of the laws of physics. This question
goes back to Ptolemy who argues in his treatise {\it On
dimensionality} that no more than three spatial dimensions are
possible in Nature. In the $18^{\rm th}$ century, Kant realizes the
deep connection between the inverse square law of gravitation and the
existence of three spatial dimensions. Interestingly, he regards the
three spatial dimensions as a consequence of Newton's inverse square
law rather than the converse. In the twentieth century, Ehrenfest
\cite{ehrenfest}, in a paper called ``In what way does it become
manifest in the fundamental laws of physics that space has three
dimensions?'' argues that planetary orbits, atoms and molecules would
be unstable in a space of dimension $D\ge 4$. This idea has been
followed more recently by Gurevich \& Mostepanenko \cite{gm} who argue
that if the universe is made of metagalaxies with various number of
dimensions, atomic matter and life are possible only in
$3$-dimensional space. Other investigations on dimensionality are
reviewed in the paper of Barrow
\cite{barrow}. We have found that the relativistic self-gravitating 
Fermi gas at $T=0$ (a white dwarf star) possesses a rich structure as
a function of the dimension of space. We have exhibited several
characteristic dimensions $D=1$, $D=2$, $D=3$, $D=(3+\sqrt{17})/2$, $D=4$ and $D=2(1+\sqrt{2})$. For $D<3$, there exists
stable configurations for any value of the mass. For $D=3$, the
sequence of equilibrium configurations is stable but there exists a
maximum mass (Chandrasekhar's limit) above which there is no
equilibrium state. For $3<D<4$, the sequence of equilibrium
configurations is stable for classical white dwarf stars but it
becomes unstable for relativistic white dwarf stars with high density
after the turning point of mass. Therefore, the dimension $D=3$ is
special because it is the largest dimension at which the sequence of
equilibrium configurations is stable all the way long; for $D>3$, a
turning point of mass appears so that ultra-relativistic white dwarf
stars become unstable. Therefore, the dimension $D=3$ is {\it
marginal} in that respect. Finally, for $D\ge 4$, the whole sequence
of equilibrium configurations is unstable.  Therefore, as already
noted in our previous papers \cite{lang,fermiD,aaantonov}, the dimension $D=4$
is critical because at that dimension quantum mechanics cannot
stabilize matter against gravitational collapse, even in the classical
regime, contrary to the situation in $D=3$ \cite{fowler,ht,pt}.  Interestingly,
this result is similar to that of Ehrenfest \cite{ehrenfest} although
it applies to white dwarf stars instead of atoms.

Our exact description of $D$-dimensional white dwarf stars based on
Chandrasekhar's seminal paper \cite{chandra3} shows that relativistic
white dwarf stars become unstable in $D>3$ and that classical white
dwarf stars become unstable in $D\ge 4$.  Therefore, for $D\ge 4$, a
self-gravitating Fermi gas forms a black hole or evaporates (see
Appendix
\ref{sec_stab}). These conclusions have also been reached by
Bechhoefer \& Chabrier
\cite{chabrier} on the basis of simple dimensional analysis.  This
suggests that a $D\ge 4$ universe is not viable (see also Appendix
\ref{sec_fourd}) and gives insight why our universe in apparently
three dimensional.  We note that extra-dimensions can appear at the
micro-scale, an idea originating from Kaluza-Klein theory. This idea
took a renaissance in modern theories of grand unification which are
formulated in higher-dimensional spaces \footnote{It is usually
assumed that these extra-dimensions appear at very small scales of the
order of the Planck length ($10^{-33}$ cm) and it is a theoretical
challenge to explain why only three dimensions are expanded while the
others are compact. We note that a universe with $D>3$ dimensions
would be very different from ours since compact objects like white
dwarfs or neutron stars would be unstable and replaced by black
holes. In fact, {\it all the matter would collapse to a single point
in a finite time} (see Appendix \ref{sec_fourd}).  This observation
may be connected to the fact that only three dimensions are
extended. }.  Our approach shows that already at a simple level, the
coupling between Newton's equations (gravitation), Fermi-Dirac
statistics (quantum mechanics) and special relativity reveals a rich
structure as a function of $D$. In this respect, it is interesting to
note that the critical masses (\ref{class10}) (\ref{ur12}) and radii
(\ref{class11}) (\ref{ur9}) that we have found occur for simple {\it
integer} dimensions $D=1,2,3$ and $4$, which was not granted a priori.

It is interesting to develop a parallel between the mass-radius
relation $M(R)$ of white dwarf stars and the caloric curve $T(E)$
giving the temperature as a function of the energy in the
thermodynamics of self-gravitating systems
\cite{lbw,katz,paddy,iso}. In this analogy, the Chandrasekhar mass in
$D=3$ is the counterpart of the critical temperature for isothermal
systems in $D=2$ (in both cases, the equilibrium density profile is a
Dirac peak containing all the mass) \footnote{There exists a formal
analogy between the Chandrasekhar limiting mass
$M_{Chandra}=5.76M_{\odot}/\mu^{2}$ for relativistic white dwarf stars
in $D=3$ (corresponding to a polytrope $n=n_{3}=3$), the critical mass
$M_{c}=8\pi$ of bacterial populations described by the Keller-Segel
model in $D=2$ (corresponding to $n=n_{3}=+\infty$) and the critical
mass $M_{c}=4k_{B}T/Gm$ or critical temperature $k_{B}T_{c}=GMm/4$ of
self-gravitating isothermal systems in $D=2$ (corresponding to
$n=n_{3}=+\infty$)
\cite{masstemp}. At these critical values $M=M_{c}$ or $T=T_{c}$, the system forms a Dirac peak. This analogy, sketched in Sec. 8 of {\tt astro-ph/0604012v1}, is addressed specifically in \cite{masscrit}.}. On the other hand, for $D>3$, the mass-radius
relation for white dwarf stars exhibits turning points, and even a
spiraling behavior for $D>{1\over 2}(3+\sqrt{17})$, which is similar
to the spiraling behavior of the caloric curve for isothermal systems
in $D=3$. In this analogy, the maximum mass, corresponding to a
critical value of the central density which parameterizes the series
of equilibria, is the counterpart of the Antonov energy (in the
microcanonical ensemble) \cite{lbw} or of the Emden temperature (in
the canonical ensemble) \cite{iso}.  The series of equilibria becomes
unstable after this turning point.  In addition, there is no
equilibrium state above this maximum mass, or below the minimum energy
or minimum temperature in the thermodynamical problem. In that case,
the system is expected to undergo gravitational collapse.

As a last comment (notified by the referee), it should be emphasized
that the conclusions reached in this paper concerning dimensionality
implicitly assume that the laws of physics that we know remain the
same in a universe of arbitrary dimension $D$. This is of course not
granted at all. There may be a new cosmological theory, a new theory
of star formation and stellar evolution in higher dimensions. We also
emphasize that our approach does not take into account general relativistic
effects that can sensibly modify the results
\cite{chandragen}.

{\it Acknowledgements:} I am grateful to E. Blackman and G. Chabrier
for pointing out their works after a first version of this paper was
placed on {\tt astro-ph/0603250}. Special thanks are due to N. Barros for
interesting discussions.

\appendix

\section{The equation of state}
\label{sec_exp}

We provide here the explicit expression of the function $f(x)$ defined by Eq. (\ref{es9}) for different dimensions of space $D=1,2,3$ and $4$, respectively:
\begin{eqnarray}
f(x)=4x(1+x^{2})^{1/2}-4\ {\sinh}^{-1}x,
\label{exp1}
\end{eqnarray}
\begin{eqnarray}
f(x)={16\over 3}+{8\over 3}(x^{2}-2)(1+x^{2})^{1/2},
\label{exp2}
\end{eqnarray}
\begin{eqnarray}
f(x)=x(2x^{2}-3)(1+x^{2})^{1/2}+3\ {\sinh}^{-1}x, 
\label{exp3}
\end{eqnarray}
\begin{eqnarray}
f(x)=-{64\over 15}+{8\over 15}(3x^{4}-4x^{2}+8)(1+x^{2})^{1/2}.
\label{exp4}
\end{eqnarray}

\section{Stability criteria for polytropic spheres in $D$ dimensions}
\label{sec_stab}

We generalize in $D$ dimensions the usual stability
criteria for polytropic gaseous spheres, and apply them to classical
and ultra-relativistic white dwarf stars.

\subsection{The Euler-Poisson system}
\label{sec_ep}

Let us consider the Euler-Poisson system  \cite{bt}:
\begin{eqnarray}
\label{stab1} {\partial\rho\over\partial t}+\nabla\cdot (\rho{\bf u})=0,
\end{eqnarray}
\begin{eqnarray}
\label{stab2} {\partial {\bf u}\over\partial t}+({\bf u}\cdot
\nabla){\bf u}= -{1\over\rho}\nabla P-\nabla\Phi,
\end{eqnarray}
\begin{eqnarray}
\label{stab3} \Delta\Phi=S_{D}G\rho,
\end{eqnarray}
describing the dynamical evolution of a barotropic gas with an equation
of state $P=P(\rho)$.  The Euler-Poisson system conserves the mass $M$ and the  energy \cite{aaantonov}:
\begin{equation}
{\cal W}=\int\rho\int^{\rho}{P(\rho')\over\rho'^{2}}d\rho' d{\bf r}+{1\over
2}\int \rho\Phi d{\bf r}+{1\over 2}\int \rho {\bf u}^{2}d{\bf r}. \label{wkre}
\end{equation}
In the following, we shall essentially consider the case of a
polytropic equation of state $P(\rho)=K\rho^{\gamma}$ where
$\gamma=1+1/n$. For $\gamma\neq 1$, the energy functional ${\cal W}$ is given by
\begin{equation}
{\cal W}={K\over \gamma-1}\int \rho^{\gamma}d{\bf r}+{1\over 2}\int \rho\Phi d{\bf r}+{1\over 2}\int \rho {\bf u}^{2}d{\bf r}. \label{qw1}
\end{equation}
It can be rewritten
\begin{equation}
{\cal W}={1\over \gamma-1}\int P d{\bf r}+{1\over 2}\int \rho\Phi d{\bf r}+{1\over 2}\int \rho {\bf u}^{2}d{\bf r}. \label{qw2}
\end{equation}
For an isothermal equation of state $P=\rho k_{B}T/m$, corresponding
to $\gamma\rightarrow 1$ or $n\rightarrow +\infty$, we have
\begin{equation}
{\cal W}=k_{B}T\int \frac{\rho}{m}\ln\frac{\rho}{m} d{\bf r}+{1\over 2}\int \rho\Phi d{\bf r}+{1\over 2}\int \rho {\bf u}^{2}d{\bf r}. \label{qw3}
\end{equation}
For a general polytropic equation of state
$P=K\rho^{\gamma}$ with arbitrary $\gamma$, it is convenient to write
the energy of the gas in the form
\begin{equation}
{\cal W}={K\over \gamma-1}\int (\rho^{\gamma}-\rho)d{\bf r}+{1\over 2}\int \rho\Phi d{\bf r}+{1\over 2}\int \rho {\bf u}^{2}d{\bf r}. \label{qw4}
\end{equation}
We have added a constant term $-{K\over \gamma-1}\int \rho d{\bf r}$
proportional to the total mass (which is a conserved quantity) so as
to recover the energy (\ref{qw3}) of an isothermal gas in the limit
$\gamma\rightarrow 1$ \cite{cstsallis}.

\subsection{The eigenvalue equation}
\label{sec_eigen}

We first consider the linear dynamical stability of a polytropic star
described by the Euler-Poisson system and generalize the approach
developed in Chavanis
\cite{iso,relativity,polytropes,grand,aaantonov,lang,virial2}. We
consider a steady solution of the Euler-Poisson system satisfying
${\bf u}={\bf 0}$ and the condition of hydrostatic balance $\nabla
P+\rho\nabla\Phi={\bf 0}$. For a polytropic equation of state
$P=K\rho^{1+{1}/{n}}$, the equilibrium density profile is solution
of the $D$-dimensional Lane-Emden equation \cite{lang}. We consider
{\it complete polytropes} such that the density vanishes at a finite
radius $R$. For $n\neq n_{3}$, there exists a unique
steady state  for any mass $M$. The mass-radius
relation is \cite{lang}:
\begin{equation}
M^{(n-1)/n}R^{(D-2)(n_{3}-n)/n}=\frac{K(1+n)}{GS^{1/n}_{D}}\omega_{n}^{(n-1)/n}, \label{dim13}
\end{equation}
where 
\begin{equation}
\omega_{n}=-\xi_{1}^{(n+1)/(n-1)}\theta_{n}'(\xi_{1}), \label{dim15}
\end{equation}
is a constant given in terms of the solution $\theta_{n}(\xi)$ of the
Lane-Emden equation of index $n$ in $D$ dimensions. For the critical
index $n=n_{3}$, steady state solutions exist only for a unique value
of the mass
\begin{equation}
M_{c}=\left\lbrack \frac{K(1+n)}{GS^{1/n}_{D}}\right\rbrack^{n/(n-1)}\omega_{n}. \label{dim13b}
\end{equation}
These solutions have the same mass $M_{c}$ but an arbitrary radius
$R$. We already anticipate that the index $n=n_{3}$ will correspond to
a case of marginal stability separating stable and unstable solutions.

Linearizing the equations of motion
(\ref{stab1})-(\ref{stab3}) around a stationary solution in
hydrostatic balance and writing the perturbation in the form
$\delta\rho\sim e^{\lambda t}$, we obtain after some calculations
\cite{iso,polytropes,virial2} the eigenvalue equation
\begin{eqnarray}
{ d\over d r}\biggl ({P'(\rho)\over S_{D}\rho r^{D-1}}{dq\over dr}\biggr )+{Gq\over r^{D-1}}={\lambda^{2}\over S_{D}\rho r^{D-1}}q,
\label{stab4}
\end{eqnarray}
where we have restricted ourselves to spherically symmetric
perturbations and defined $q(r)=\int_{0}^{r}\delta\rho
S_{D}r^{D-1}dr$. For a polytropic gas with an equation of state
$P=K\rho^{\gamma}$, the foregoing equation becomes \cite{polytropes}:
\begin{eqnarray}
K\gamma { d\over d r}\biggl ({\rho^{\gamma-2}\over S_{D} r^{D-1}}{dq\over dr}\biggr )+{Gq\over r^{D-1}}={\lambda^{2}\over S_{D}\rho r^{D-1}}q.
\label{stab5}
\end{eqnarray}
The polytrope is stable if $\lambda^{2}<0$ (yielding oscillatory modes
with pulsation $\omega=\sqrt{-\lambda^{2}}$) and unstable if
$\lambda^{2}>0$ (yielding exponentially growing modes with growth rate
$\gamma=\sqrt{\lambda^{2}}$).  Considering the point of marginal
stability ($\lambda=0$) and introducing the Emden variables
\cite{chandrabook,polytropes,lang}, Eq. (\ref{stab5}) reduces to
\begin{eqnarray}
{ d\over d r}\biggl ({\theta^{1-n}\over \xi^{D-1}}{dF\over d\xi}\biggr )+{nF\over \xi^{D-1}}=0.
\label{stab6}
\end{eqnarray}
This equation has the exact solution \cite{polytropes,lang}:
\begin{eqnarray}
F(\xi)=c_{1}\biggl\lbrack \xi^{D}\theta^{n}+{(D-2)n-D\over n-1}\xi^{D-1}\theta'\biggr\rbrack.
\label{stab7}
\end{eqnarray}
The point of marginal stability is then determined by the boundary
conditions. One can show \cite{polytropes,virial2} that the velocity
perturbation is given by $\delta u=-\lambda q/(S_{D} \rho
r^{D-1})$. Therefore, if the density of the configuration vanishes at
$r=R$, one must have $q(R)=0$ to avoid unphysical divergences. Thus,
if $\xi_1$ denotes the value of the normalized radius $R$ of the star
\cite{chandrabook,polytropes,lang} such that $\theta(\xi_{1})=0$, the
natural boundary condition for the eigenvalue equation (\ref{stab5})
is $F(\xi_{1})=0$.  Substituting this condition in Eq. (\ref{stab7}),
we obtain the critical index
\begin{eqnarray}
n={D\over D-2}\equiv n_{3}, \label{stab8}
\end{eqnarray}
corresponding to a marginally stable gaseous polytrope
($\lambda=0$). Our method gives the form (\ref{stab7}) of the neutral
perturbation $\delta\rho$ at the critical index $n=n_{3}$. We conclude
that the infinite family of steady state solutions with equal mass $M_{c}$
and arbitrary radius $R$ that exists for $n=n_{3}$ is marginally
stable.

We now consider the nonlinear dynamical stability problem.  Since the
Euler-Poisson system (\ref{stab1})-(\ref{stab3}) conserves the mass
$M$ and the energy ${\cal W}$, a maximum or a minimum of the energy
functional ${\cal W}[\rho,{\bf u}]$ at fixed mass $M[\rho]=M$
determines a steady state of the Euler-Poisson system that is formally
nonlinearly dynamically stable
\cite{aaantonov}. Because of the presence of the kinetic term 
$\Theta[{\bf u}]=(1/2)\int \rho {\bf u}^{2}d{\bf r}$, the functional
${\cal W}[\rho,{\bf u}]$ has no absolute maximum. Thus, we need to
investigate the possible existence of a minimum of ${\cal W}[\rho,{\bf
u}]$ at fixed mass $M[\rho]=M$. A barotropic star that minimizes the
energy functional ${\cal W}$ at fixed mass $M$ is nonlinearly
dynamically stable with respect to the Euler-Poisson system. The
cancellation of the first order variations $\delta {\cal
W}-\alpha\delta M=0$, where $\alpha$ is a Lagrange multiplier, yields
${\bf u}={\bf 0}$ and the condition of hydrostatic balance $\nabla
P+\rho\nabla\Phi={\bf 0}$.  Then, the condition of nonlinear dynamical
stability is
\begin{eqnarray}
\delta^{2}{\cal W}=\int {P'(\rho)\over 2\rho}(\delta\rho)^{2}d{\bf r}+{1\over 2}\int\delta\rho\delta\Phi d{\bf r}\ge 0, \label{fdf}
\end{eqnarray}
for all perturbations that conserve mass, i.e. $\int\delta\rho d{\bf r}=0$. After some calculations \cite{aaantonov}, this can be put in the quadratic form
\begin{eqnarray}
\delta^{2}{\cal W}=-{1\over 2}\int_{0}^{R}\left\lbrack {d\over dr}\left ({P'(\rho)\over S_{D}\rho r^{D-1}}{dq\over dr}\right )+{Gq\over r^{D-1}}\right\rbrack q \ dr. \nonumber\\
\label{fdfq}
\end{eqnarray}  
We are led therefore to consider the eigenvalue problem
\begin{eqnarray}
\left\lbrack {d\over dr}\left ({P'(\rho)\over S_{D}\rho r^{D-1}}{d\over dr}\right )+{G\over r^{D-1}}\right\rbrack q_{\lambda}(r)=\lambda q_{\lambda}(r). \label{fdf2}
\end{eqnarray}  
If all the eigenvalues $\lambda$ are negative, then $\delta^{2}{\cal
W}>0$ and the configuration is a minimum of ${\cal W}$ at fixed
mass. This implies that it is nonlinearly dynamically stable. If at
least one eigenvalue $\lambda$ is positive, the configuration is a
saddle point of ${\cal W}$ and the star is dynamically unstable. The
marginal case is when the largest eigenvalue $\lambda$ is equal to
zero. Now, for $\lambda=0$, Eqs. (\ref{fdf2}) and (\ref{stab4})
coincide. This implies that the conditions of linear and nonlinear
dynamical stability are the same. In the case of polytropic stars, the
case of marginal stability corresponds to the
critical index $n=n_{3}$. At that index, the equilibrium
configurations with mass $M_{c}$ and radius $R$ {\it all} have the
same value of energy ${\cal W}=0$ (see Eq. (\ref{stab22}) later).
This is therefore a very degenerate situation.

The nonlinear dynamical stability of gaseous polytropes can also be
investigated by plotting the series of equilibria $M(\rho_{0})$ (mass
vs central density) of box-confined configurations and using the
turning point argument of Poincar\'e
\cite{polytropes,lang,aaantonov}. We can thus determine whether the last point on the series of equilibria, which corresponds to a {\it complete polytrope} 
whose density vanishes precisely at the box radius, is stable or
not. For $D\le 2$, there is no turning point of mass (we restrict
ourselves to $n\ge 0$), implying that the gaseous polytropes are
always stable. For $D>2$, a turning point of mass $M(\rho_{0})$
appears precisely for $n=n_3$ (see Fig. 5 of \cite{lang}). This method
shows that, for $D>2$, complete polytropes with $n<n_3$ are
nonlinearly dynamically stable (they are minima of ${\cal W}$ at fixed
mass $M$) while complete polytropes with $n>n_3$ are dynamically
unstable (they are saddle points of ${\cal W}$ at fixed mass
$M$). Complete polytropes with $n=n_3$ and $M=M_{c}$ are marginally
stable.

\subsection{The Ledoux criterion}
\label{sec_ledoux}

We can also investigate the linear dynamical stability of gaseous
polytropic spheres by using the method introduced by Eddington \cite{eddington2} and Ledoux \cite{ledoux}. If we introduce the radial displacement
\begin{eqnarray}
\xi(r)=-{\delta u\over \lambda r}={q\over S_{D}\rho r^{D}}\propto {\delta r\over r},
\label{stab9}
\end{eqnarray}
we can rewrite the eigenvalue equation (\ref{stab5}) in the form \cite{virial1,virial2}:
\begin{eqnarray}
{d\over dr}\biggl (P\gamma r^{D+1}{d\xi\over dr}\biggr )+r^{D}(D\gamma+2-2D){dP\over dr}\xi=\lambda^{2}\rho r^{D+1}\xi.\nonumber\\
\label{stab10}
\end{eqnarray}
This is the Eddington equation of pulsations, which has been written
here in the form of a Sturm-Liouville problem. It must be supplemented
by the boundary conditions
\begin{eqnarray}
\delta r=\xi r=0, \qquad {\rm in }\quad r=0,
\label{bc1}
\end{eqnarray}
\begin{eqnarray}
dP=\lambda\gamma P\left (D\xi+r\frac{d\xi}{dr}\right )=0,\qquad {\rm in }\quad r=R,
\label{bc2}
\end{eqnarray}
where $dP/dt=\partial\delta P/\partial t+\delta u dP/dr$ is the Lagrangian derivative of the pressure. Since $P=0$ at the surface of the star, it is sufficient to demand that $\xi$ and $d\xi/dr$ be finite in $r=R$.  Multiplying Eq. (\ref{stab10}) by $\xi$ and integrating
between $0$ and $R$, we obtain
\begin{eqnarray}
\lambda^{2}\int_{0}^{R}\rho r^{D+1}\xi^{2}dr
=-\int_{0}^{R} dr \biggl\lbrace P\gamma r^{D+1}\biggl ({d\xi\over dr}\biggr )^{2}
\nonumber\\
-\xi^{2}r^{D}(D\gamma+2-2D){dP\over dr}\biggr\rbrace.
\label{stab11}
\end{eqnarray}
The system is linearly dynamically stable if $\lambda^{2}< 0$ and
unstable otherwise.  Since $dP/dr<0$, a sufficient condition of
stability is $D\gamma+2-2D> 0$, i.e.
\begin{eqnarray}
\gamma> \gamma_{4/3}\equiv {2(D-1)\over D}, \qquad {1\over n}>
{1\over n_{3}}\equiv {D-2\over D}. \label{stab12}
\end{eqnarray}
It can be shown furthermore (see below) that the system is unstable if
$\gamma<\gamma_{4/3}$ so that the criterion (\ref{stab12}) is a
necessary and sufficient condition of dynamical stability (the case
$\gamma=\gamma_{4/3}$ is marginal).  In terms of the index $n$, a complete
polytrope is stable with respect to the Euler-Poisson system in $D=1$
for $n\ge 0$ and for $n<-1$, in $D=2$ for $n> 0$, in $D=3$ for $0\le
n< 3$ and in $D>2$ for $0\le n< n_3$.

From the theory of Sturm-Liouville problems, it is known that
expression (\ref{stab11}), which can be written
$\lambda^{2}=I\lbrack\xi\rbrack$, forms the basis of a variational
principle. The function $\xi(r)$ which maximizes the functional
$I\lbrack\xi\rbrack$ is the fundamental eigenfunction and the maximum
value of this functional gives the fundamental eigenvalue
$\lambda^{2}$. Furthermore, any trial function under-estimates the
value of $\lambda^{2}$ so this variational principle may prove the
existence of instability but can only give approximate information
concerning stability.  As shown by Ledoux \& Pekeris \cite{ledoux}, we
can get a good approximation of the fundamental eigenvalue by taking
$\xi(r)$ to be a constant (note that $\xi={\rm Cst.}$, i.e. $\delta
r\propto r$, is the {\it exact} solution of the Sturm-Liouville
equation (\ref{stab10}) at the point of marginal stability $\lambda=0$
for a polytropic equation of state). For the trial function $\xi={\rm
Cst.}$, expression (\ref{stab11}) gives
\begin{eqnarray}
\lambda^{2}\int \rho r^{2} d{\bf r}=(D\gamma+2-2D)\int r {dP\over dr} d{\bf r}.
\label{stab13}
\end{eqnarray}
According to the condition of hydrostatic balance,
\begin{eqnarray}
{dP\over dr}=-\rho{d\Phi\over dr},
\label{stab14}
\end{eqnarray}
we have
\begin{eqnarray}
\int r {dP\over dr} d{\bf r}=-\int\rho {\bf r}\cdot \nabla\Phi d{\bf r},
\label{stab15}
\end{eqnarray}
where we recognize the Virial
\begin{eqnarray}
W_{ii}\equiv -\int\rho {\bf r}\cdot \nabla\Phi d{\bf r}.
\label{stab16}
\end{eqnarray}
Inserting 
\begin{eqnarray}
\nabla\Phi=G\int
\rho({\bf r}'){{\bf r}-{\bf r}'\over |{\bf r}-{\bf r}'|^{D}}d{\bf
r}',
\label{stab17b}
\end{eqnarray}
in Eq. (\ref{stab16}), interchanging the dummy variables ${\bf r}$
and ${\bf r}'$ and taking the half-sum of the resulting expressions, we get
\begin{eqnarray}
W_{ii}=-{G\over 2}\int {\rho({\bf r})\rho({\bf r}')\over |{\bf r}-{\bf r}'|^{D-2}}d{\bf r}d{\bf r}'.
\label{stab16bis}
\end{eqnarray}
Therefore, the Virial can be written
\begin{eqnarray}
W_{ii}=(D-2)W, \qquad (D\neq 2),
\label{stab18}
\end{eqnarray}
\begin{eqnarray}
W_{ii}=-{GM^{2}\over 2}, \qquad (D=2),
\label{stab19}
\end{eqnarray}
where 
\begin{eqnarray}
W=\frac{1}{2}\int \rho\Phi d{\bf r},
\label{stab19b}
\end{eqnarray}
is the potential energy and
\begin{eqnarray}
\Phi=-\frac{G}{D-2}\int{\rho({\bf r}')\over |{\bf r}-{\bf
r}'|^{D-2}}d{\bf r}',
\label{stab19c}
\end{eqnarray}
is the gravitational potential.  Therefore, we can rewrite
Eq. (\ref{stab13}) in the form
\begin{eqnarray}
\lambda^{2}=(D\gamma+2-2D){W_{ii}\over I},
\label{stab17}
\end{eqnarray}
where
\begin{eqnarray}
I=\int\rho r^{2}d{\bf r},
\label{stab17a}
\end{eqnarray}
is the moment of inertia.  Since $W_{ii}<0$, we conclude that the
system is unstable if $D\gamma+2-2D< 0$, which completes the proof
above. On the other hand, Eq. (\ref{stab17}) provides an estimate of
the pulsation period $\omega=\sqrt{-\lambda^{2}}$ when the system is
stable. This is the $D$-dimensional generalization of the Ledoux
stability criterion.

Note finally that, for a spherically symmetric system, using the Gauss theorem
\begin{eqnarray}
\nabla\Phi=\frac{GM(r)}{r^{D-1}}{\bf e}_{r},
\label{stab19d}
\end{eqnarray}
we get
\begin{eqnarray}
W_{ii}=-S_{D}G\int\rho(r)M(r)r \, dr=-\int\frac{GM(r)}{r^{D-2}}dM(r).\nonumber\\
\label{stab19e}
\end{eqnarray}
This expression may be useful to calculate the potential energy.

\subsection{Virial theorem and  Poincar\'e argument}

The Virial theorem associated with the barotropic Euler-Poisson system (\ref{stab1})-(\ref{stab3}) is \cite{bt,virial2}:
\begin{equation}
{1\over 2}{d^{2}I\over dt^{2}}=2\Theta+\Pi+W_{ii}, \label{vir1}
\end{equation}
where $\Theta=(1/2)\int \rho {\bf u}^{2}d{\bf r}$ is the kinetic energy of
the macroscopic motion and $\Pi=D\int Pd{\bf r}$. For a polytropic
equation of state, the energy functional (\ref{wkre}), which is a
conserved quantity, can be written
\begin{equation}
{\cal W}=U+W+\Theta, \label{vir1b}
\end{equation}
where
\begin{equation}
U=\frac{1}{\gamma-1}\int Pd{\bf r}, \label{vir1c}
\end{equation}
is the internal energy.  We note that $\Pi=D(\gamma-1)U=(D/n)U$ for a
polytrope. At equilibrium, $\ddot I=\Theta=0$, we get $(D/n)U+W_{ii}=0$
and ${\cal W}=U+W$. For $D=2$, this implies that $U=nGM^{2}/4$. For $D\neq
2$, this implies that
\begin{equation}
{\cal W}=\left (1-{n\over n_{3}}\right )W, \label{stab22}
\end{equation}
where we recall that
\begin{equation}
W=-{G\over 2(D-2)}\int {\rho({\bf r})\rho({\bf r}')\over |{\bf
r}-{\bf r}'|^{D-2}}d{\bf r}d{\bf r}'. \label{stab23}
\end{equation}
For $D>2$ and $n\ge 0$, the system is dynamically stable if ${\cal
W}<0$ and unstable otherwise (Poincar\'e argument
\cite{chandrabook}). Since $W<0$, we find from Eq. (\ref{stab22}) that the polytropic star is stable if and
only if $n<n_{3}$. At the point of marginal stability $n=n_{3}$, where
several steady configurations exist with the same mass $M_{c}$ given
by Eq. (\ref{dim13b}) and an arbitrary radius $R$, the energy of these
configurations is ${\cal W}=0$ for all $R$.

Using ${\cal W}=(n/D)\Pi+W+\Theta$ and eliminating the pressure term in
Eq. (\ref{vir1}), the Virial theorem for $D\neq 2$ can be put in the
form
\begin{equation}
{1\over 2}{d^{2}I\over dt^{2}}=\left (2-{D\over n}\right )\Theta+{D\over n}{\cal W}+\left (D-2-{D\over n}\right )W. \label{vir2}
\end{equation}
Alternatively, eliminating the kinetic energy in
Eq. (\ref{vir1}),  the
Virial theorem for $D\neq 2$ can be written
\begin{equation}
{1\over 2}{d^{2}I\over dt^{2}}=2{\cal W}+\left ({D\over n}-2\right )U+(D-4)W. \label{vir2b}
\end{equation}
For $n=n_{3/2}$, corresponding to classical white dwarf stars, the Virial theorem becomes
\begin{equation}
{1\over 2}{d^{2}I\over dt^{2}}=2{\cal W}+(D-4)W. \label{vir2c}
\end{equation} 
For $n=n_{3}'$, corresponding to relativistic white dwarf stars, we have
\begin{equation}
{1\over 2}{d^{2}I\over dt^{2}}=\Theta+{\cal W}+(D-3)W. \label{vir2d}
\end{equation} 
Finally, considering the polytropic index $n=n_{3}$ of marginal
stability, the Virial theorem (\ref{vir2}) takes the form
\begin{equation}
{1\over 2}{d^{2}I\over dt^{2}}=(4-D)\Theta+(D-2){\cal W}. \label{vir3}
\end{equation} 
If we consider the dimension $D=4$, it reduces to
\begin{equation}
{d^{2}I\over dt^{2}}=4{\cal W}. \label{vir4}
\end{equation} 
This equation describes the case of classical white dwarf stars at the
critical dimension $D=4$ where $n_{3/2}=n_{3}=2$. It can be integrated
into
\begin{equation}
I(t)=2{\cal W}t^{2}+C_{1}t+C_{2}. \label{vir5}
\end{equation} 
For ${\cal W}>0$, we find that $I(t)\rightarrow +\infty$ for
$t\rightarrow +\infty$ so that the system evaporates. Alternatively,
for ${\cal W}<0$, we find that $I(t)\rightarrow 0$ in a finite time,
so that the system collapses and forms a Dirac peak in a finite
time. If we consider the dimension $D=3$, Eq. (\ref{vir3}) reduces to
\begin{equation}
\frac{1}{2}{d^{2}I\over dt^{2}}=\Theta+{\cal W}. \label{vir4n}
\end{equation} 
This equation describes the case of relativistic  white dwarf stars at the
critical dimension $D=3$ where $n_{3}'=n_{3}=3$. For ${\cal W}>0$, we find that $I(t)\rightarrow +\infty$ for $t\rightarrow +\infty$ indicating that the system evaporates.

\subsection{Dimensional analysis}

Finally, we show that the instability criterion for polytropic stars
can be obtained from simple dimensional analysis. We shall approximate the
internal energy (\ref{vir1c}) and the potential energy (\ref{stab19b}) by
\begin{equation}
U={K\zeta\over \gamma-1} \left ({M\over R^{D}}\right )^{\gamma}R^{D}, \label{dim1a}
\end{equation}
\begin{equation}
W=-{\nu\over D-2}{GM^{2}\over R^{D-2}}, \label{dim1b}
\end{equation}
where $M$ is the total mass of the configuration and $R$ its radius
($\nu$ and $\zeta$ are dimensionless parameters). For homogeneous
spheres, the values of $\zeta$ and $\nu$ are given by
Eq. (\ref{dim1c}).  With these expressions, the energy functional
(\ref{qw1}) becomes
\begin{equation}
{\cal W}={K\zeta\over \gamma-1} \left ({M\over R^{D}}\right )^{\gamma}R^{D}
-{\nu\over D-2}{GM^{2}\over R^{D-2}}+\Theta. \label{dim1}
\end{equation}
We now need to minimize this functional with respect to $R$
for a given mass $M$ (a minimum necessarily requires $\Theta=0$). We first
look for the existence of critical points (extrema). The cancellation
of the first order variations
\begin{equation}
{d{\cal W}\over dR}=0=-K\zeta DM^{\gamma}R^{D(1-\gamma)-1}+G\nu M^{2}R^{1-D}, \label{dim2}
\end{equation}
yields the mass-radius relation
\begin{equation}
M^{(n-1)/n}R^{(D-2)(n_{3}-n)/n}={K\zeta D\over G\nu}. \label{dim3}
\end{equation}
This relation determines the radius $R$ of the star as a function of
its mass $M$. This expression is consistent with the exact mass-radius
relation (\ref{dim13}) deduced from the Lane-Emden equation (of
course, our simple approach can only model compact density profiles
corresponding to $n<n_{5}$ for $D>2$)
\cite{lang}. For $n\neq n_{3}$, there is one, and only one, extremum
of ${\cal W}(R)$ for each mass $M$.  In order to have a true minimum,
we need to impose
\begin{equation}
{d^{2}{\cal W}\over dR^{2}}=-G\nu DM^{2}R^{-D}\left ({1\over
n_{3}}-{1\over n}\right )> 0. \label{dim4}
\end{equation}
Therefore, the system is stable if
\begin{equation}
{1\over n}> {1\over n_{3}}, \label{dim5}
\end{equation}
and unstable otherwise. This simple dimensional analysis returns the
exact stability criterion (\ref{stab12}).

We can be a little more precise. For $1/n>1/n_{3}$, the functional
${\cal W}(R)$ has a global minimum reached for a finite, and non-zero,
value of $R$. This solution is stable.  For $1/n<1/n_3$, the
functional ${\cal W}(R)$ has an unstable global maximum: for $D<2$,
${\cal W}\rightarrow -\infty$ when $R\rightarrow +\infty$
(evaporation) and ${\cal W}\rightarrow 0$ when $R\rightarrow 0$
(collapse); for $D=2$, ${\cal W}\rightarrow -\infty$ when
$R\rightarrow 0$ and $R\rightarrow +\infty$; for $D>2$ and $n<0$,
${\cal W}\rightarrow -\infty$ when $R\rightarrow 0$ and $R\rightarrow
+\infty$; for $D>2$ and $n>n_{3}$, ${\cal W}\rightarrow -\infty$ when
$R\rightarrow 0$ and ${\cal W}\rightarrow 0$ when $R\rightarrow
+\infty$. Since we have constructed a particular configuration
(homogeneous sphere) which makes the energy ${\cal W}$ diverge to
$-\infty$, the above arguments prove that the exact functional ${\cal
W}[\rho,{\bf u}]$ given by Eq. (\ref{wkre}) has no absolute minimum at
fixed mass for $1/n<1/n_3$. Since the functional ${\cal W}[\rho,{\bf
u}]$ has only one critical point (cancelling the first variations) at
fixed mass, we conclude that, for $1/n<1/n_3$, this critical point is
a saddle point. Therefore, there is no stable steady state of
polytropic spheres for $1/n<1/n_3$. For the critical index $n=n_{3}$,
the relation (\ref{dim3}) shows the existence of a critical mass
\begin{equation}
M_{c}=\left ( {K\zeta D\over G\nu}\right )^{D/2}. \label{dim3b}
\end{equation}
The functional (\ref{dim1}) can be rewritten
\begin{equation}
{\cal W}=K\zeta\frac{D}{D-2}\frac{M^{2(D-1)/D}}{R^{D-2}}\left \lbrack 1-\left (\frac{M}{M_{c}}\right )^{2/D}\right\rbrack+\Theta. \label{aw1}
\end{equation}
For $M=M_{c}$, ${\cal W}(R)=0$ for all $R$ (at equilibrium
$\Theta=0$). This determines an infinite family of solutions with the
same mass $M_{c}$ and different radii. These solutions have the same
energy and are marginally stable. This returns the result of the exact
model where the configurations are solution of the Lane-Emden equation
(see Sec. \ref{sec_eigen}). For $M<M_{c}$, the function ${\cal W}(R)$
is monotonically decreasing with $R$ (so the system tends to
evaporate): for $D<2$, ${\cal W}(R)$ goes from $0$ to $-\infty$ and
for $D>2$, ${\cal W}(R)$ goes from $+\infty$ to $0$. For $M>M_{c}$,
the function ${\cal W}(R)$ is monotonically increasing with $R$ (so
the system tends to collapse): for $D<2$, ${\cal W}(R)$ goes from $0$
to $+\infty$ and for $D>2$, ${\cal W}(R)$ goes from $-\infty$ to $0$.

We can obtain a simple dynamical model by using the Virial theorem
(\ref{vir1}). At equilibrium ($\ddot I=\Theta=0$), we have
\begin{equation}
\frac{D}{n}U+(D-2)W=0. \label{dim6}
\end{equation}
Inserting the expressions (\ref{dim1a}) and (\ref{dim1b}) in
Eq. (\ref{dim6}), we recover the mass-radius relation
(\ref{dim3}). For $n\neq n_{3}$ and for any given mass $M$, there is
only one steady state. Its radius $R_{0}$ given by Eq. (\ref{dim3})
and its energy ${\cal W}_{0}$ is given by Eq. (\ref{stab22}). It
corresponds to the extremum value of ${\cal W}(R)$.  Estimating the
moment of inertia by
\begin{equation}
I=\alpha MR^{2}, \label{dim7}
\end{equation}
and inserting the expressions (\ref{dim1a}) and (\ref{dim1b}) in
Eq. (\ref{vir2b}), we obtain
\begin{eqnarray}
\frac{1}{2}\alpha M \frac{d^{2}R^{2}}{dt^{2}}=2{\cal W}+(D-2n)K\zeta M^{1+1/n}R^{-D/n}\nonumber\\
-\frac{\nu(D-4)}{D-2}\frac{GM^{2}}{R^{D-2}}.\qquad\label{dim8}
\end{eqnarray}
This equation determines the evolution of the radius of the star for a
fixed mass $M$ and a fixed energy ${\cal W}$. The evolution of the
kinetic energy is then given by $\Theta={\cal W}-U-W$. The solution of
Eq. (\ref{dim8}) depends on two control parameters $M$ and ${\cal W}$
and on the initial condition $R(0)$ and ${\dot R}(0)$. We shall consider
the case where ${\cal W}$ is equal to the value ${\cal W}_{0}$
corresponding to the steady state, such that the r.h.s. of the above
equation is equal to zero at equilibrium. Then, we can rewrite
Eq. (\ref{dim8}) as
\begin{eqnarray}
\frac{1}{2}\alpha M \frac{d^{2}R^{2}}{dt^{2}}=(D-2n)K\zeta M^{1+1/n}(R^{-D/n}-R_{0}^{-D/n})\nonumber\\
-\frac{\nu(D-4)}{D-2}\left( \frac{GM^{2}}{R^{D-2}}-\frac{GM^{2}}{R_{0}^{D-2}}\right ),\qquad
\label{dim9}
\end{eqnarray}
where $R_{0}$ is the radius of the star at equilibrium. Since the
dynamics is non-dissipative \footnote{We could simply account for
dissipative effects by introducing a friction force $-\xi {\bf u}$ in
the Euler equation (\ref{stab2}) \cite{virial2}. In that case
$\dot{\cal W}=-\xi\int\rho {\bf u}^{2}d{\bf r}\le 0$, so that the
system evolves so as to minimize the functional ${\cal W}[\rho,{\bf u}]$
at fixed mass.}, the system does not evolve towards the minimum of
${\cal W}(R)$ (unless $R=R_{0}$ initially). The system can either
oscillate around the minimum (stable) or evolve away from it
(unstable). Considering small perturbations around equilibrium,
writing $R=R_{0}(1+\epsilon)$ with $\epsilon\ll 1$, and linearizing
the foregoing equation, we obtain
\begin{eqnarray}
\frac{d^{2}\epsilon}{dt^{2}}+(D\gamma+2-2D)\frac{\nu GM}{\alpha R_{0}^{D}}\epsilon=0,
\label{dim10}
\end{eqnarray} 
where we have used the mass-radius relation (\ref{dim3}) to simplify
the expression. This is the equation for a harmonic oscillator with pulsation
\begin{eqnarray}
\omega^{2}=(D\gamma+2-2D)\frac{\nu GM}{\alpha R^{D}}.
\label{dim11}
\end{eqnarray} 
The star is stable if $\omega^{2}>0$ and unstable otherwise. This
returns the exact stability criterion (\ref{stab12}). Furthermore,
using Eqs. (\ref{dim1b}) and (\ref{dim7}), the pulsation can be
rewritten in the form
\begin{eqnarray}
\omega^{2}=-(D\gamma+2-2D)\frac{W_{ii}}{I},
\label{dim12}
\end{eqnarray} 
which exactly coincides with the Ledoux formula
(\ref{stab17}). Therefore, our simple dimensional model allows to
obtain a lot of interesting results. The case of arbitrary
perturbations around equilibrium will be considered in a future work.
Finally, we note that for $n=n_{3}$, Eq. (\ref{dim8}) becomes
\begin{eqnarray}
\frac{1}{2}\alpha M \frac{d^{2}R^{2}}{dt^{2}}=2{\cal W}+K\zeta \frac{D(D-4)}{D-2}\frac{M^{2(D-1)/D}}{ R^{D-2}}\nonumber\\
\times \left\lbrack 1-\left (\frac{M}{M_{c}}\right )^{2/D}\right\rbrack,
\label{dim8fg}
\end{eqnarray}
and Eq. (\ref{dim9}) corresponds to ${\cal W}=0$ and $M=M_{c}$
yielding $d^{2}R^{2}/dt^{2}=0$. Hence, $R^{2}=C_{1}t+C_{2}$
corresponding to a marginal evolution.

In our dimensional analysis, the constants $\zeta$, $\nu$ and $\alpha$
are dimensionless parameters that could be chosen to fit the exact
results at best. Alternatively, we can try to obtain quantitative
predictions by calculating their values for homogeneous
spheres. This yields
\begin{equation}
\zeta=\left (\frac{D}{S_{D}}\right )^{1/n},\qquad \nu=\alpha=\frac{D}{D+2}. \label{dim1c}
\end{equation}
Note that the potential energy of a homogeneous sphere in $D$
dimensions can be easily calculated from Eq. (\ref{stab19e}). On the
other hand, in $D=2$, a direct calculation gives
$W=(1/2)GM^{2}\ln(R/L)-GM^{2}/8$ where $L$ is a reference radius where
$\Phi(L)=0$. All the results given above pass to the limit
$D\rightarrow 2$ provided that we take $\nu=1/2$. Using
Eq. (\ref{dim1c}), we find that the approximate value of the pulsation
(\ref{dim11}) becomes
\begin{eqnarray}
\omega^{2}=(D\gamma+2-2D)\frac{GM}{R^{D}}.
\label{dim11ht}
\end{eqnarray} 
On the other hand, if we compare the approximate mass-radius relation
(\ref{dim3}) with the exact mass-radius relation (\ref{dim13}),
we obtain an estimate of the constant $\omega_{n}$ in the form
\begin{equation}
\omega^{approx.}_{n}=\left (\frac{D+2}{n+1}\right )^{n/(n-1)}D^{1/(n-1)}. \label{dim14}
\end{equation}
This has to be compared with the exact value (\ref{dim15})
given in terms of the solution $\theta_{n}(\xi)$ of the Emden equation
of index $n$ in $D$ dimensions \cite{lang}. Let us consider the case
$D=3$. For $n=3/2$, corresponding to classical white dwarf stars, we
find $\omega_{3/2}^{approx.}=72$ instead of the exact value
$\omega_{3/2}=132.3843...$. For $n=3$, corresponding to relativistic
white dwarf stars, we find $\omega_{3}^{approx.}=2.42...$ instead of
the exact value $\omega_{3/2}=2.01824...$ This suggests that the
homogeneous star model will provide a fair description of
relativisitic white dwarf stars and a poorer description of
classical white dwarf stars. We shall come back to these different
issues (static and dynamics) in a future work.

\section{Energy functionals}
\label{sec_ef}

In this Appendix, we show that the condition of thermodynamical stability in the canonical ensemble is equivalent to the condition of nonlinear dynamical stability with respect to the barotropic Euler-Poisson system. We apply this result to white dwarf stars. 

\subsection{Energy of a barotropic gas}
\label{sec_efa}

We consider a barotropic gas with an equation of state $P=P(\rho)$
described by the Euler-Poisson system \cite{bt}. We introduce the energy
functional
\begin{equation}
{\cal W}=\int\rho\int^{\rho}{P(\rho')\over\rho'^{2}}d\rho' d{\bf r}+{1\over 2}\int \rho\Phi d{\bf r}+{1\over 2}\int \rho {\bf u}^{2}d{\bf r}.
\label{ef1}
\end{equation}
The first term ${\cal W}_{1}$ is the work $-P(\rho)d(1/\rho)$ done in
compressing the system from infinite dilution, the second term $W$ is
the gravitational energy and the third term $\Theta$ is the kinetic
energy associated with the mean motion.  For a gas in local
thermodynamic equilibrium, the equation of state is $P=P(\rho,T)$ or
$P=P(\rho,s)$ and the first law of thermodynamics can be written
$d(u/\rho)=-Pd(1/\rho)+Td(s/\rho)$ where $s$ is the entropy density
and $u$ is the density of internal energy. For a gas without
interaction (apart from the long-range gravitational attraction), the
internal energy is equal to the kinetic energy. There are two
important cases where the gas is barotropic. The first case is when
$Td(s/\rho)=0$. This concerns either adiabatic (or isentropic) fluids
so that $d(s/\rho)=0$ or fluids at zero temperature so that
$T=0$. When $Td(s/\rho)=0$, the first law of thermodynamics reduces to
$d(u/\rho)=-Pd(1/\rho)$. It can be integrated into $u=\rho\int^{\rho}
[P(\rho')/\rho'^{2}]d\rho'$. Then, the work ${\cal W}_{1}$ done
by the pressure force (first term in Eq. (\ref{ef1})) coincides with
the internal energy $U$ of the gas. In that case, we get ${\cal
W}_{1}=U$ and the energy functional (\ref{ef1}) can be written
\begin{equation}
{\cal W}=U+W+\Theta=E.\label{ef2}
\end{equation}
Thus, at $T=0$ or for an adiabatic evolution, the total energy of the
gas $E$ is conserved by the Euler-Poisson system (since ${\cal W}$ is
conserved).  Alternatively, for an isothermal gas $dT=0$, the first
law of thermodynamics $d(u/\rho)=-Pd(1/\rho)+Td(s/\rho)$ can be
written $d((u-Ts)/\rho)=-Pd(1/\rho)$. It can be integrated into
$u-Ts=\rho\int^{\rho}
[P(\rho')/\rho'^{2}]d\rho'$.  Therefore, the work ${\cal W}_{1}$ done
by the pressure force (first term in Eq. (\ref{ef1})) coincides with
the free energy $U-TS$ of the gas. In that case, we get ${\cal
W}_{1}=U-TS$ and the energy functional (\ref{ef1}) can be written
\begin{equation}
{\cal W}=U-TS+W+\Theta=E-TS=F.\label{ef3}
\end{equation}
Thus, for an isothermal evolution, the free energy of the system $F$
is conserved by the Euler-Poisson system (since ${\cal W}$ is
conserved). At $T=0$, we recover the conservation of the energy $E$.

Let us apply these results to white dwarf stars.  We can view a white
dwarf star as a barotropic gas described by an equation of state
$P=P(\rho)$. According to the discussion of Appendix \ref{sec_eigen}
(see also \cite{aaantonov}), it is nonlinearly dynamically stable with
respect to the Euler-Poisson system if it is a minimum of the energy
functional ${\cal W}$ at fixed mass. If a minimum exists, it is
necessary that ${\bf u}=0$. As a result, a steady state of the
Euler-Poisson system is nonlinearly dynamically stable if, and only
if, it is a minimum of $\tilde{\cal W}[\rho]={\cal W}[\rho,{\bf
u}]-\Theta[{\bf u}]$ at fixed mass $M$. In conclusion, the condition of
nonlinear dynamical stability can be written
\begin{equation}
{\rm Min} \quad \lbrace \tilde{\cal W}[\rho]\quad |\quad M[\rho]=M\rbrace.
\label{ef4}
\end{equation}
For a white dwarf star at zero temperature ($T=0$), according to
Eqs. (\ref{ef2}) and (\ref{ef3}), the functional $\tilde{\cal W}$
reduces to the energy $E=U+W$, where $U$ is the kinetic energy.

\subsection{Free energy of self-gravitating fermions}
\label{sec_efb}

We can also view a white dwarf star as a gas of self-gravitating
relativistic fermions at statistical equilibrium in the canonical
ensemble. It is thermodynamically stable if, and only if, it is a minimum of free
energy $F$ at fixed mass $M$.  The free energy is given by
\begin{eqnarray}
F=E-TS=\int f\epsilon(p)d{\bf p}d{\bf r}+{1\over 2}\int \rho\Phi d{\bf r}\nonumber\\
+{T\eta_{0}\over m}\int \biggl \lbrace {f\over\eta_{0}}\ln {f\over\eta_{0}}+\biggl (1-{f\over\eta_{0}}\biggr )\ln\biggl (1-{f\over\eta_{0}}\biggr )\biggr\rbrace d{\bf r}d{\bf p},\nonumber\\
\label{ef5}
\end{eqnarray}
where $\eta_{0}=2/h^{D}$ is the maximum value of the distribution
function fixed by the Pauli exclusion principle. For a white dwarf
star at zero temperature ($T=0$), the free energy $F$ reduces to the
energy $E=U+W$. The condition of thermodynamical stability in the
canonical ensemble can be written
\begin{equation}
{\rm Min} \quad \lbrace {F}[f]\quad |\quad M[f]=M\rbrace.
\label{ef6}
\end{equation}
The critical points of free energy at fixed mass, determined by the
variational principle $\delta F-\alpha\delta M=0$, correspond to the
relativistic mean-field Fermi-Dirac distribution
\begin{eqnarray}
f=\frac{\eta_{0}}{1+e^{\beta\lbrack \epsilon(p)+\mu H\Phi({\bf r})+\lambda_{0}\rbrack}},
\label{ef7}
\end{eqnarray}
where $\lambda_{0}$ is a Lagrange multiplier (related to $\alpha$)
determined by the mass $M$ and $\epsilon(p)$ is given by
Eq. (\ref{es3}).  Using Eqs. (\ref{es1}), (\ref{es4}) and (\ref{ef7}),
the density and the pressure are of the form $\rho=\rho(\mu H\Phi({\bf
r})+\lambda_{0})$ and $P=P(\mu H\Phi({\bf
r})+\lambda_{0})$. Eliminating $\mu H\Phi({\bf r})+\lambda_{0}$
between these two expressions, we find that $P=P(\rho)$ so that the
gas is barotropic. The equation of state is parameterized by $T$ and
is fully determined by the entropic functional in Eq. (\ref{ef5}),
which is here the Fermi-Dirac entropy. At $T=0$ we obtain the explicit
relations of Sec. \ref{sec_es}.  Furthermore, the condition that
$f({\bf r},{\bf p})$ is a function of the energy $e=\epsilon(p)+\mu
H\Phi({\bf r})$ implies that the corresponding barotropic gas is at
hydrostatic equilibrium. Indeed, using Eq. (\ref{es5}),
\begin{eqnarray}
\nabla P=\frac{1}{D}\int \frac{\partial f}{\partial {\bf r}}p\epsilon'(p)d{\bf p}=\mu H\nabla\Phi\frac{1}{D}\int f'(e)p\epsilon'(p)d{\bf p}\nonumber\\
=\mu H\nabla\Phi\frac{1}{D}\int \left ({\bf p}\cdot \frac{\partial f}{\partial {\bf p}}\right )d{\bf p}=-\mu H\nabla\Phi\int f d{\bf p},\nonumber\\
 \label{ef7bg}
\end{eqnarray}
so that
\begin{eqnarray}
\nabla P=-\rho\nabla\Phi.
 \label{ef7bgc}
\end{eqnarray}

The relativistic mean-field Fermi-Dirac distribution (\ref{ef7}) is
just a critical point of free energy at fixed mass $M$.  To determine
whether it corresponds to a true minimum of free energy, we can
proceed in two steps. We first minimize $F[f]$ for a fixed density
profile $\rho({\bf r})$. Since the potential energy $W[\rho]$ and the
mass $M[\rho]$ are entirely determined by the density profile, this is
equivalent to minimizing $U[f]-TS[f]$ at fixed $\rho({\bf r})$, where
$U$ is the kinetic energy. This gives a distribution
\begin{eqnarray}
\tilde{f}=\frac{\eta_{0}}{1+e^{\beta\lbrack \epsilon(p)+\lambda({\bf r})\rbrack}},
\label{ef8}
\end{eqnarray}
where $\lambda({\bf r})$ is a local Lagrange multiplier determined by
the density $\rho({\bf r})$, using $\rho=\mu H\int f d{\bf p}$. Since
$\delta^{2}(U-TS)\ge 0$, the distribution (\ref{ef8}) is a true
minimum of $F[f]$ at fixed $\rho({\bf r})$. Substituting the optimal
distribution function (\ref{ef8}) in Eq. (\ref{ef5}), we obtain a
functional $\tilde{F}[\rho]\equiv F[\tilde{f}]$ of the density
$\rho({\bf r})$. Using Eqs. (\ref{es1}), (\ref{es4}) and (\ref{ef8}),
we note that the density and the pressure are of the form $\rho({\bf
r})=\rho(\lambda({\bf r}))$ and $P({\bf r})=P(\lambda({\bf
r}))$. Eliminating $\lambda({\bf r})$ between these expressions, we
find that $P=P(\rho)$ where the equation of state is the same as the
one determined from the Fermi-Dirac distribution (\ref{ef7}) at
equilibrium. Now, we can show that
\begin{eqnarray}
\tilde{F}[\rho]=\tilde{\cal W}[\rho],
\label{ef9}
\end{eqnarray}
where $\tilde {\cal W}[\rho]$ is the functional defined in
Sec. \ref{sec_efa}. The relation (\ref{ef9}) can be obtained by an
explicit calculation which extends the proof given in Appendix B of
\cite{aaantonov} for classical particles (this will be shown in 
a future work; see also particular cases in Sec. \ref{sec_efc}) but it
also results from a straightforward argument. We note that the
distribution function (\ref{ef8}) corresponds to a condition of local
thermodynamical equilibrium with uniform temperature and zero average
velocity. Thus, it locally satisfies the first law of thermodynamic
$d(u/\rho)=-Pd(1/\rho)+Td(s/\rho)$ with $dT=0$. Integrating this
relation like in Sec. \ref{sec_efa}, we find that
$U[\rho]-TS[\rho]={\cal W}_{1}[\rho]$ where $U[\rho]=U[\tilde{f}]$ and
$S[\rho]=S[\tilde{f}]$. This directly yields the identity (\ref{ef9}).

We can now conclude that $F[f]$ has a minimum $f_{*}({\bf r},{\bf p})$
at fixed mass $M$ if, and only if, $\tilde{F}[\rho]=\tilde {\cal
W}[\rho]$ has a minimum $\rho_{*}({\bf r})$ at fixed mass $M$. In that
case, $f_{*}({\bf r},{\bf p})$ is given by Eq. (\ref{ef8}) where
$\lambda_{*}({\bf r})$ is determined by $\rho_{*}({\bf r})$, writing
$\rho_{*}=\mu H\int f_{*} d{\bf p}$. Therefore, a system is
thermodynamically stable in the canonical ensemble if, and only if,
the corresponding barotropic gas with the same equilibrium
distribution is nonlinearly dynamically stable with respect to the
barotropic Euler-Poisson system. Said differently, a system that
minimizes the functional (\ref{ef9}) is (i) thermodynamically stable
in the canonical ensemble and (ii) nonlinearly dynamically stable with
respect to the barotropic Euler-Poisson system. This result applies to
 white dwarf stars at arbitrary temperature.

\subsection{Application to white dwarf stars at $T=0$}
\label{sec_efc}

Although the above results are general, it may be useful to explicitly
compute the functionals (\ref{ef1}) and (\ref{ef5}) for white dwarf
stars at $T=0$ and check the relation (\ref{ef9}).  If we view a white dwarf
star as a barotropic gas described by an equation of state $P(\rho)$,
its energy $\tilde{\cal W}$ can be written 
\begin{equation}
\tilde{\cal W}=\int\rho\Gamma(\rho)d{\bf r}+{1\over 2}\int \rho\Phi d{\bf r},
\label{ef10}
\end{equation}
with
\begin{equation}
\Gamma(\rho)=\int^{\rho}{P(\rho')\over\rho'^{2}}d\rho'.
\label{ef11}
\end{equation}
A white dwarf star is nonlinearly dynamically stable with respect to
the Euler-Poisson system if it is a minimum of $\tilde{\cal W}$ at
fixed mass $M$.   At $T=0$, the equation of state is given by
Eq. (\ref{es7}). In the classical limit (polytrope $n_{3/2}=D/2$) we
have
\begin{equation}
\tilde{\cal W}={DK_{1}\over 2}\int \rho^{(D+2)/ D}d{\bf r}+{1\over 2}\int \rho\Phi d{\bf r},
\label{ef12}
\end{equation}
and in the ultra-relativisitic limit (polytrope $n_{3}'=D$) we have
\begin{equation}
\tilde{\cal W}={DK_{2}}\int \rho^{(D+1)/ D}d{\bf r}+{1\over 2}\int \rho\Phi d{\bf r}.
\label{ef13}
\end{equation}
In the general case, using Eq. (\ref{es7}), the function (\ref{ef11})
can be written
\begin{eqnarray}
\Gamma(\rho)=\frac{A_{2}D}{B}\int^{x}\frac{f(x')}{x'^{D+1}}dx'.
\label{ef14}
\end{eqnarray}
Integrating by parts and using Eqs. (\ref{es8}) and (\ref{es9}), we find after
straightforward calculations that
\begin{eqnarray}
\frac{\mu H}{mc^{2}}\Gamma(\rho)=\sqrt{1+x^{2}}-\frac{1}{x^{D}}\int_{0}^{x}\frac{t^{D+1}}{(1+t^{2})^{1/2}}dt.\nonumber\\
\label{ef15}
\end{eqnarray}

Alternatively, we can view a white dwarf star as a gas of
self-gravitating relativistic fermions at statistical equilibrium in
the canonical ensemble.  At $T=0$, its free energy $F=E-TS$ coincides with
its energy $E$.  Using Eq. (\ref{ef5}), it is given by
\begin{eqnarray}
F=E=\int \frac{\rho}{\mu H}\kappa(\rho)d{\bf r}+{1\over 2}\int \rho\Phi d{\bf r},
\label{ef16}
\end{eqnarray}
with 
\begin{eqnarray}
\frac{\rho}{\mu H}\kappa(\rho)=\int f\epsilon(p)d{\bf p}.
\label{ef17}
\end{eqnarray}
A white dwarf star at $T=0$ is thermodynamically stable if
it is a minimum of the energy $E$ at fixed mass $M$. In the classical
limit, $\epsilon={p^{2}/2m}$ and $P={1\over D}\int (f/m)
p^{2}d{\bf p}$ so $\rho\kappa/\mu H=(D/2)P$. Therefore, the free energy can be written
\begin{eqnarray}
\tilde{F}={D\over 2}\int P d{\bf r}+{1\over 2}\int \rho\Phi d{\bf r}.
\label{ef18}
\end{eqnarray}
Using Eq. (\ref{class1}), this is equivalent to Eq. (\ref{ef12}).
In the ultra-relativistic limit, $\epsilon=pc$ and $P={1\over D}\int fpc d{\bf p}$ so $\rho\kappa/\mu H=D P$. Therefore, the free energy can be written
\begin{eqnarray}
\tilde{F}={D}\int P d{\bf r}+{1\over 2}\int \rho\Phi d{\bf r}.
\label{ef19}
\end{eqnarray}
Using Eq. (\ref{ur1}), this is equivalent to Eq. (\ref{ef13}). In the
general case, we have
\begin{eqnarray}
\kappa(\rho)=\frac{2S_{D}}{nh^{D}}\int_{0}^{p_{0}}\epsilon(p)p^{D-1}dp.
\label{ef20}
\end{eqnarray}
Using Eq. (\ref{es3}), it can be put in the form
\begin{eqnarray}
\frac{\kappa(\rho)}{mc^{2}}=\frac{D}{x^{D}}\int_{0}^{x} \sqrt{1+t^{2}}t^{D-1}dt.
\label{ef21}
\end{eqnarray}
Now, it is straightforward to check that the two expressions in the r.h.s. of Eqs. (\ref{ef15}) and (\ref{ef21}) are equal so that $\Gamma(\rho)={\kappa(\rho)}/{\mu H}$ implying the relation (\ref{ef9}).

Finally, if we consider a classical isothermal self-gravitating gas at temperature $T$ with an equation of state $P=\rho k_{B}T/m$, its free energy (\ref{ef5}) can be written
\begin{equation}
\tilde{F}=\frac{D}{2}Nk_{B}T+k_{B}T\int \frac{\rho}{m}\ln\frac{\rho}{m} d{\bf r}+{1\over 2}\int \rho\Phi d{\bf r}. \label{ef22}
\end{equation}
Comparing with Eq. (\ref{qw3}), we find that relation (\ref{ef9}) is
indeed satisfied.  Therefore, a classical isothermal self-gravitating
gas that minimizes the functional (\ref{qw3}) or (\ref{ef22}) at fixed
mass is (i) thermodynamically stable in the canonical ensemble and
(ii) nonlinearly dynamically stable with respect to the barotropic
Euler-Poisson system. We had already made this observation in
\cite{iso}. In \cite{aaantonov}, this result was extended to an 
arbitrary form of entropic functional, including the Fermi-Dirac
entropy, for non-relativistic systems. The present paper shows that,
due to relation (\ref{ef9}), the equivalence between nonlinear
dynamical stability with respect to the barotropic Euler-Poisson
system and thermodynamical stability in the canonical ensemble is
general.

\section{Non viability of a $D\ge 4$ universe}
\label{sec_fourd}

If we consider, in a $D$-dimensional universe, a Hamiltonian system of
self-gravitating classical point masses whose dynamics is described by
the Newton equations
\begin{equation}
\ddot{\bf r}_{\alpha}=\sum_{\beta\neq\alpha}{Gm({\bf r}_{\beta}-{\bf r}_{\alpha})\over |{\bf r}_{\beta}-{\bf r}_{\alpha}|^{D}}, \label{fourd1}
\end{equation}
the scalar Virial theorem reads \cite{cras}:
\begin{equation}
{1\over 2}\ddot I=2K+W_{ii}, \label{fourd2}
\end{equation}
where $K$ is the kinetic energy and $W_{ii}$ the trace of the
potential energy tensor for the $N$-body system \cite{cras}.  For
$D\neq 2$, using Eq. (\ref{stab18}) and introducing the total energy
$E=K+W$, which is a constant of the motion for an isolated system, we
have \cite{cras}:
\begin{equation}
{1\over 2}\ddot I=2K+(D-2)W=2E+(D-4)W. \label{fourd3}
\end{equation}
We note that the dimension $D=4$ is critical. In that case, ${\ddot
I}=4E$ which yields after integration $I=2Et^{2}+C_{1}t+C_{2}$. For
$E>0$, $I\rightarrow +\infty$ indicating that the system
evaporates. For $E<0$, $I$ goes to zero in a finite time, indicating
that the system forms a Dirac peak (``black hole'') in a finite
time. More generally, for $D\ge 4$, since $(D-4)W\le 0$, we have $I\le
2Et^{2}+C_{1}t+C_{2}$ so that the system forms a Dirac peak in a
finite time if $E<0$. Therefore, self-gravitating systems with $E<0$
are not stable in a space of dimension $D\ge 4$. The study of the
present paper indicates that this observation remains true if
quantum (Pauli exclusion principle for fermions) and relativistic
effects are taken into account. In this sense, a universe with $D\ge 4$ is
not viable.

\end{document}